\pdfoutput=1
\documentclass[nonacm,sigplan]{acmart}

\usepackage{soul}
\usepackage{xcolor}
\usepackage{tikz}
\usepackage{amsmath}

\usepackage[euler]{textgreek}
\usepackage[most]{tcolorbox}
\usepackage{listings}
\usepackage{cleveref}
\usepackage[normalem]{ulem}
\usepackage{enumitem}
\usepackage{subcaption}

\newcommand{\ufont}{\ttfamily}
\newcommand{\mfont}{\ttfamily}

\newcommand{\ourspectre}{\textmu{}Spec\-tre}
\newcommand{\Ourspectre}{\ourspectre{}}

\newcommand{\ourmitigation}{\textmu{}SLH}
\newcommand{\Ourmitigation}{\ourmitigation{}}
\newcommand{\ourslh}{\ourmitigation{}}

\newcommand{\microop}{micro-op}
\newcommand{\microops}{\microop{}s}
\newcommand{\macroop}{macro-op}

\newcommand{\msrom}{MSROM}

\newcommand{\uspectre}{\ourspectre{}}

\newcommand{\hlc}[2][yellow]{{\sethlcolor{#1}\hl{#2}}}

\newcommand{\lstseq}[1]{\textbf{#1}}
\newcommand{\lstbr}[1]{\textcolor{red}{\textbf{#1}}}
\newcommand{\lsttrans}[1]{\textcolor{gray}{\textbf{#1}}}
\newcommand{\lstbad}[1]{\hlc[pink]{#1}}
\newcommand{\lstnew}[1]{\textcolor{Green4}{\textbf{#1}}}
\newcommand{\lstsec}[1]{\hlc[yellow]{#1}}

\newcommand{\ucode}{\textmu{}code}
\newcommand{\ubranches}{\textmu{}branches}
\newcommand{\ubranch}{\textmu{}branch}

\newcommand{\uops}{\textmu{}ops}
\newcommand{\upc}{\textmu{}pc}

\newcommand{\meb}{\uspectre{}-MEB}
\newcommand{\mvi}{\uspectre{}-MVI}
\newcommand{\mil}{\uspectre{}-MIL}

\definecolor{Green4}{RGB}{46, 126, 27}

\lstdefinestyle{example}{
  basicstyle=\ttfamily\footnotesize,
  numbers=left,
}

\lstdefinestyle{ucode}{
    basicstyle=\ufont\footnotesize,
    numbers=left,
}

\lstdefinestyle{mcode}{
    basicstyle=\mfont\footnotesize,
    numbers=left,
}

\begin{document}

\date{}

\title{Analyzing and Exploiting Branch Mispredictions in Microcode}

\author{Nicholas Mosier}
\email{nmosier@stanford.edu}
\orcid{0000-0003-3705-8161}
\affiliation{%
    \institution{Stanford University}
    \city{Stanford}
    \country{USA}
}

\author{Hamed Nemati}
\email{hnnemati@kth.se}
\orcid{0000-0001-9251-3679}
\affiliation{%
    \institution{KTH Royal Institute of Technology}
    \city{Stockholm}
    \country{Sweden}
}

\author{John C. Mitchell}
\email{jcm@stanford.edu}
\orcid{0000-0002-4065-3327}
\affiliation{%
    \institution{Stanford University}
    \city{Stanford}
    \country{USA}
}

\author{Caroline Trippel}
\email{trippel@stanford.edu}
\orcid{0000-0002-5776-1121}
\affiliation{%
    \institution{Stanford University}
    \city{Stanford}
    \country{USA}
}

\begin{abstract}
We present \textit{\ourspectre{}}, a new class of transient execution attacks that exploit \textit{microcode branch mispredictions} to transiently leak sensitive data.
We find that many long-known and recently-discovered transient execution attacks, which were previously categorized as Spectre or Meltdown variants, are actually instances of \ourspectre{} on some Intel microarchitectures.
Based on our observations, we discover multiple new \ourspectre{} attacks and present a defense against \ourspectre{} vulnerabilities, called \textit{\ourmitigation{}}.

\end{abstract}

\maketitle
\pagestyle{plain}

\section{Introduction}
\label{sec:intro}
Modern processors typically implement CISC ISAs, like x86, by translating each architectural instruction (\textit{\macroop{}}) into a sequence of one or more micro-operations (\textit{\microop{}}s)
at decode time, which are then issued to the execution unit and executed independently.
Translation of simple macro-ops, which generate a single \microop{} (e.g., an x86 load \texttt{mov eax, [rsp]}) or even a few \microops{} (e.g., three \microops{} for \texttt{call [rax]} on x86~\cite{uops.info}), 
can be incorporated directly into the decode logic itself.
However, complex macro-ops (like \texttt{cpuid}) require so many micro-ops to implement that their implementations must be stored in a special micro-instruction memory, called the \textit{microsequencer ROM} (\msrom{})~\cite{dead-uops}.

The \msrom{} is a collection of \textit{microprograms}, each of which implements a single macro-op (or a family of related macro-ops).
Micro-ops stored in the \msrom{} are collectively called \textit{microcode}.

Like architectural programs, microprograms may contain branches, including conditional branches.
These conditional \textit{microcode branches} (\textit{\ubranches{}}) transfer control to another microcode address within the same microprogram, conditioned on some predicate computed at runtime.
As with conditional macro-op branches, the CPU \textit{predicts} the direction of a \ubranch{} before its predicate has been computed
and \textit{speculatively executes} \microops{} along the predicted microcode control-flow path.
If the microprogram exits, speculative execution continues at the next macro-op.
Recent research~\cite{hackers-ucodedisasm} has discovered that on Intel's Goldmont microarchitecture~\cite{intel_optimization_manual}, conditional \ubranches{} are statically predicted not-taken.
Our experiments show this also is the case on other Intel microarchitectures (Goldmont Plus, Haswell, Kaby Lake, and Golden Cove).

\subsection{This Paper}
In this paper, \textbf{our key observation} is that, like conditional (macro-op) branch mispredictions,
conditional \ubranch{} mispredictions can create transient execution windows spanning many macro-ops, during which macro-ops may leak sensitive data.

Based on this observation, \textbf{our first contribution} is the discovery of
a new class of transient execution attacks, called \textit{\ourspectre{}}, which exploit conditional \ubranch{} mispredictions to transiently leak sensitive data.

Importantly, \Ourspectre{} differs from the two established classes of transient execution attacks: Spectre~\cite{spectre, spectre-rsb, spectre-stl, retbleed, spectre-psf, spectre-btc} (including Phantom~\cite{phantom}) and Meltdown~\cite{meltdown, foreshadow, lvi, mds-fallout, mds-ridl, mds-zombieload, lazyfp}.
Unlike Spectre and Phantom, which exploit mispredictions at the macro-op level,  \ourspectre{} exploits mispredictions at the micro-op level within the CPU's internal \ucode{}.
Unlike Meltdown, \ourspectre{} does not involve delayed fault handling~\cite{canella-systematic}.

However, \ourspectre{} attacks can \textit{masquerade} as Spectre and Meltdown %
attacks at the macro-op level,
depending on what the mispredicted \ubranch{} is guarding. 
Indeed, we find that many long-known transient execution attacks (e.g., Rogue System Register Read~\cite{intel-rogue-system-register-read}) as well as more recently discovered attacks (e.g., Zero Dividend Injection~\cite{revizor++}), previously thought to be %
Spectre or Meltdown variants, are in fact instances of \ourspectre{} on some Intel microarchitectures.

After defining \uspectre{} attacks, as \textbf{our second contribution},
we report \textit{multiple new \ourspectre{} vulnerabilities} that we found during our evaluation of the microcode of the Intel Goldmont microarchitecture. For example, we discover
\textit{Out-of-Bounds Performance Counter Read} (\ourspectre{}-MEB-PCIDX),
which we exploit to leak the contents of a microarchitectural buffer that is architecturally inaccessible at \textit{all} privilege levels (\S\ref{sec:new:pcidx}).

To mitigate \ourspectre{} attacks, \textbf{our third contribution} is the
\textit{\ourmitigation{} defense}, inspired by speculative load hardening (SLH, \cite{llvm-slh}), which inserts a data dependency on microcode branch predicates to force
architectural registers that may hold unauthorized/incorrect data to return zero along mispredicted microcode paths.
\Ourmitigation{} can be deployed on existing Intel CPUs via microcode updates to fix all discovered \ourspectre{} vulnerabilities.

\paragraph{Contributions.}
We summarize our contributions as follows.
\begin{itemize}[leftmargin=*]
    \item \textit{Characterization of \ubranch{} execution behaviors (\S\ref{sec:ubranches}): }
    We characterize the execution and (mis)prediction behaviors of microcode branches
    on a selection of Intel Atom (Goldmont, Goldmont Plus, Gracemont)
    and Core (Haswell, Kaby Lake, Golden Cove)
    microarchitectures.
    To our knowledge, we are the first to document the following: (1) the size, in macro-ops, of transient execution windows that \ubranch{} mispredictions introduce;
    (2) that \ubranch{} mispredictions are detected out-of-order but corrected in-order at retirement; and
    (3) that the targets of indirect \ubranches{} are not predicted.

    \item \textit{\uspectre{} attacks (\S\ref{sec:uspectre}):}
    We present \textit{\uspectre{}}, a novel class of transient execution attacks that exploit conditional \ubranch{} mispredictions
    to leak sensitive architectural or microarchitectural state without authorization.
    We identify \textit{three} variants of \uspectre{} attacks: \textit{Microcode Exception Bypass} (\textbf{\uspectre{}-MEB}), \textit{Microcode Value Injection} (\textbf{\uspectre{}-MVI}), and \textit{Microcode-Internal Leakage} (\textbf{\uspectre{}-MIL}).
    Finally, we extend the taxonomy of Canella et al.~\cite{canella-systematic} to include \uspectre{} and its variants.

    \item \textit{Reclassifying Known Transient Execution Attacks as \uspectre{} (\S\ref{sec:existing}):}
    We demonstrate how a number of known Spectre and Meltdown attacks---specifically, Rogue System Register Read~\cite{intel-rogue-system-register-read}, Meltdown-BND~\cite{canella-systematic}, Zero Dividend Injection~\cite{revizor++}, and String Comparison Overrun~\cite{revizor++}---are in fact instances of \uspectre{}.
    When possible, we provide the actual microprograms that implement the vulnerable instructions and explain how the observed vulnerabilities arise due to mispredicted conditional microcode branches.

    \item \textit{Disclosing New \uspectre{} Vulnerabilities (\S\ref{sec:new}):}
    We disclose two new \uspectre{} vulnerabilities---\textit{Overflow Check Bypass} (\textbf{\uspectre{}-MEB-OF}) and \textit{Out-of-Bounds Performance Counter Read} (\textbf{\uspectre{}-MEB-PCIDX})---that impact Intel Goldmont processors and possibly others.
    The latter, to our knowledge, is the first transient execution attack capable of leaking %
    the contents of an internal, architecturally inaccessible microarchitectural buffer that, to our knowledge, does not directly correspond to any architectural state. 

    \item \textit{Mitigating \uspectre{} with \ourmitigation{} (\S\ref{sec:uslh}):}
    We present \ourmitigation{}, a \uspectre{} defense inspired by speculative load hardening~\cite{llvm-slh}.
    \ourmitigation{} uses \textit{conditional select} micro-ops to conditionally zero out sensitive data following a microcode branch misprediction.
    \ourmitigation{} can be deployed via microcode updates to 
    mitigate most vulnerabilities discussed in this paper,
    including Rogue System Register Read and \uspectre{}-MEB-PCIDX (\S\ref{sec:uslh}). %

\end{itemize}

\textbf{Responsible Disclosure.} 
We disclosed an earlier version of this report to Intel in June 2024 and this version in January 2025. Intel shared our report with its industry partners, but has not released a CVE or any microcode updates in response to our findings.

\section{Background}

In this section, we give background on transient execution attacks, of which \uspectre{} is one, and Intel microcode.

\label{sec:background}

\subsection{Hardware Side Channel Attacks}
\label{sec:background:hw-side-channels}
\label{sec:background:transmitter}
In a hardware side-channel attack, a transmit instruction (or \textit{transmitter}) in the victim program
modulates (changes the state of) a channel (a hardware resource) in an operand-dependent manner,
and a receiver (the attacker) observes the channel modulation to infer the operand value~\cite{dawg:kiriansky}.
Hardware resources that can form channels include caches~\cite{Osvik:prime+probe+l1d,Yarom:flush+reload+llc13,cache_bleed}, branch predictors~\cite{branchpred_sc, Evtyushkin:BranchScope}, functional units~\cite{andrysco:subnormal,Mult_leaky}, memory ports~\cite{portsmash}, and others~\cite{yan:directories,memjam,xu:controlledchannelattacks,leaky_cauldron,tlb_bleed,drama,pandora:isca:21,augury:2022:vicarte:oakland22}.
A receiver observes a transmitter's channel modulations via their effects on non-deterministic aspects of program execution, e.g.,
execution time~\cite{bernstein_aes_attack,exectimeexploit,Osvik:prime+probe+l1d}, 
resource contention~\cite{branchpred_sc, andrysco:subnormal, portsmash, memjam, xu:controlledchannelattacks, drama, tlb_bleed}, 
and more~\cite{perf_counter_sc,powerexploit, acousticexploit, radiationexploit,KocherDPA,sok_em_side_channels}.

\subsection{Transient Execution Attacks}
\label{sec:background:meltdown}
\textit{Transient execution attacks}~\cite{canella-systematic} are hardware side channel attacks 
where the transmitter is transient (i.e., bound-to-squash~\cite{stt}).
In other words, transient execution attacks leak via (transient) transmitters sensitive architectural or microarchitectural state without authorization.

Canella et al.~\cite{canella-systematic} systematically classify all transient execution attacks known at the time of publication as either Spectre attacks~\cite{spectre, spectre-rsb, spectre-stl, amd-predictive-store-forwarding, retbleed, oleksenko:revizor, phantom, spectre-btc, spectre-sls, rage-against-the-machine-clear, inception}, which exploit control-flow or data-flow mispredictions, or Meltdown attacks~\cite{meltdown, foreshadow, lvi, mds-fallout, mds-ridl, mds-zombieload, lazyfp}, which exploit delayed fault handling.
While we generally adopt the Canella et al. classification scheme, we refine it in \S\ref{sec:uspectre:taxonomy} to include \uspectre{} and other recent attacks~\cite{spectre-btc, spectre-psf, phantom} that have been discovered since the scheme's first publication.

\subsection{Micro-ops, Microprograms, and Microcode}
\label{sec:background:msrom}
\label{sec:background:upc}
Rather than executing architectural instructions (\textit{macro-ops}) directly,
modern x86 processors first translate them into a sequence of one or more RISC-like \textit{micro-operations} (\textit{micro-ops}, or \uops{} for short) before passing them to the CPU's backend for execution~\cite{intel_optimization_manual}.
This translation process is performed at decode time.

The decoder has two units for translating macro-ops to micro-ops.
The \textit{Micro-Instruction Translation Engine} (MITE) uses hard-wired decode logic to directly translate \textit{simple} macro-ops into three or fewer micro-ops~\cite{intel_optimization_manual} (or, on Intel's recent Golden Cove microarchitecture, four).
The \textit{microsequencer} indirectly translates \textit{complex} macro-ops
(into three or more micro-ops)
by fetching their micro-op implementations---called \textit{microprograms}---from an in-core read-only memory called the \textit{microseqeuncer ROM} (MSROM).
\textit{Microcode} collectively refers to all microprograms
stored in the MSROM.
We say a macro-op is \textit{microcoded} if its microprogram resides in the MSROM, and it is thus decoded by the microsequencer, \textit{not} the MITE.
A microcoded macro-op is associated with a \textit{microcode entrypoint}, the first microcode address in its microprogram.

Because microprograms may contain many micro-ops (e.g., as many as 138 for \texttt{xsave} on Golden Cove~\cite{uops.info}), the microsequencer generally cannot fetch the entire microprogram in a single cycle.
Thus, the microsequencer maintains a \textit{microprogram counter} (\upc{}), which holds the next MSROM address from which to fetch micro-ops.
At each cycle, the microsequencer fetches up to three micro-ops from the current \upc{} (or four, on some recent Intel processors like Golden Cove)
and then updates the next \upc{} accordingly. 
Once the current microprogram ends (indicated with the \verb|SEQW UEND0| sequence word, \S\ref{sec:background:ucode-details}),
the microsequencer loads the microcode entrypoint of the next microcoded macro-op into \upc{} and proceeds fetching micro-ops from there.

Like architectural (macro-op) code, microcode may contain branches, including conditional branches, which transfer control to another microcode address by updating the \upc{}.
\textit{These conditional microcode branches (\ubranches{}) and the class of transient execution attacks they give rise to are the focus of this paper.
}
\subsubsection{Microcode on Goldmont}
Ermolov et al. dumped and reverse engineered (most of) the microcode on Intel's Goldmont efficiency core~\cite{intel-earlier-uarches} and published their results~\cite{hackers-ucodedisasm, hackers-ucodedump, hackers-writeup}.
To our knowledge, Goldmont is the \textit{only} Intel microarchitecture whose microcode is publicly available. 
Thus, we will frequently reference Goldmont microcode and microprograms throughout this paper. 

\subsubsection{Microcode on Other Cores}
While we do not have a microcode dump for Intel microarchitectures beyond the Goldmont efficiency core, we
infer that the microcode of more recent Intel efficiency cores---e.g., Goldmont Plus, Tre\-mont, and Gracemont---closely resembles that of Goldmont, 
based on how closely the values of microcode-related performance counters match up across generations.
For example, results published to uops.info~\cite{uops.info} indicate that the microprogram implementing the macro-op \texttt{rcl m16, cl} consists of 18/18/19/18 micro-ops on Goldmont / Goldmont Plus / Tremont / Gracemont efficiency cores, respectively.
In contrast, all Intel performance cores from Sandy Bridge (2011) through Golden Cove (2021) use 10 or 11 micro-ops~\cite{uops.info}.

\subsubsection{Microcode Details}\label{sec:background:ucode-details}
In this section, we discuss relevant features of the microcode on Intel Goldmont efficiency cores (and most likely all other modern Intel efficiency cores), based on the findings of Ermolov et al.~\cite{hackers-ucodedisasm, hackers-writeup}.
Please see their report for more details.

\paragraph{Microcode register file}
Micro-op register operands can encode both architectural registers
as well as non-arch\-itect\-ural scratch registers.
We define the \textit{microcode register file} to be the set of registers that are directly addressable via micro-op register operands.

Goldmont's microcode register file includes architectural integer and floating-point registers (e.g., \texttt{rax} and \texttt{xmm0}) as well as
non-architectural integer and float\-ing-point scratch registers (\texttt{tmp0}--\texttt{tmp15} and \texttt{tmm0}--\texttt{tmm15}, respectively).
Scr\-atch registers are used to hold the results of intermediate macro-op computations.

\paragraph{Arithmetic flags}
In addition to maintaining the global \texttt{RFLAGS} architectural register, 
microcode also attaches arithmetic flags to \textit{each integer register} in the microcode register file.
Upon writing to an integer register, the CPU updates its associated arithmetic flags based on the computation that produced value being written.

\paragraph{Triads}\label{sec:background:triad}
Microcode in the MSROM is organized into micro-op \textit{triads}, groups of three micro-ops.
When active, the microsequencer fetches one triad per cycle from the MSROM.
We find in \S\ref{sec:ubranches:pred:uncond-ind} that the microsequencer speculatively ignores (i.e., does not take) microcode branches that are not the last (i.e., third) in their triad, even if the branch is unconditional.

\paragraph{Sequence words}\label{sec:background:seqw}
In addition to micro-ops, the MSROM stores \textit{sequence words}, one per micro-op triad.
Sequence words encode what we call \textit{microsequencer directives}, which influence how the microsequencer fetches subsequent micro-ops. %
We write microsequencer directives as \texttt{SEQW <direct\-ive>} in our microcode listings throughout the paper.
Relevant microsequencer directives for this paper are the following:
\begin{itemize}
    \item \texttt{SEQW UEND0} marks the end of the microprogram.
    It instructs the microsequencer to stop fetching micro-ops.

    \item \texttt{SEQW LFNCEWAIT} delays the execution of all subsequent micro-ops until all \textit{loads or load-like micro-ops} (e.g., control register reads) that precede the most recent \texttt{SEQW LFNCEMARK} have executed.
    Microprograms implicitly start with a \texttt{SEQW LFNCEMARK}.

    \item \texttt{SEQW SYNCWAIT} delays the execution of all subsequent micro-ops until all \textit{instructions} that precede the most recent \texttt{SEQW SYNCMARK} have executed.
    Microprograms implicitly start with a \texttt{SEQW SYNCMARK}, based on the behaviors we have observed on Goldmont.

    \item \texttt{SEQW SYNCFULL} delays the execution of all subsequent micro-ops until all prior micro-ops have executed. 

    \item \texttt{SEQW GOTO uaddr} informs the microsequencer to redirect control to microcode address \texttt{uaddr} (i.e., $\text{\upc{}} \gets \texttt{uaddr}$).
    Unlike \ubranches{}, \texttt{SEQW GOTO} control-flow transfers are \textit{not micro-ops} and are resolved directly in the microsequencer at fetch/decode time. 
    Furthermore, \texttt{SEQW GOTO}s are never predicted (discussed in \S\ref{sec:ubranches:ucf}).
    
\end{itemize}
Ermolov et al.~\cite{hackers-writeup, hackers-ucodedisasm} are the first to document the semantics of these microsequencer directives.
Their work contains a more comprehensive list of microsequencer directives and their semantics. %

\section{Microcode Branches}
\label{sec:ubranches}
\label{sec:ubranches:cores}

In this section, we characterize the behavior of microcode branches in modern Intel CPUs.
All conclusions we make are based on our observations when studying the Intel Goldmont (Apollo Lake, Celeron N3350), Goldmont Plus (Gemini Lake, Celeron N4000), Haswell (Haswell, Core i3-4170T), and Golden Cove (Alder Lake, Core i9-12900KS) microarchitectures (platform and SKUs in parentheses).

\subsection{Microcode Control-Flow}
\label{sec:ubranches:ucf}
Based on the reverse engineering by Ermolov et al. of Intel Goldmont's microcode~\cite{hackers-ucodedisasm},
Intel microcode has \textit{two} distinct mechanisms to redirect microarchitectural control flow:
\textit{microcode sequence words} and \textit{microcode branches}.

Some microcode sequence words (\S\ref{sec:background:seqw}), like \texttt{SEQW GOTO}, are control-flow transfer directives that the microsequencer handles at decode time. %
Microcode sequence words can encode both unconditional direct control-flow transfers and conditional direct control-flow transfers that depend solely on the processor's current operating mode (e.g., the current ring, whether in a VM, etc.) but, importantly, not any runtime data (e.g., whether a flag in a control register is set).

\textit{Microcode branches}, in contrast, are micro-ops that encode control-flow transfers that \textit{may depend on data computed at runtime}.
Microcode branches grant microcode the ability to make \textit{data-dependent} decisions, which are necessary to implement complex instructions in the x86 ISA.

\begin{table*}[htbp]
    \centering
    \begin{tabular}{|l|l|p{3.4cm}|p{3.4cm}|p{4cm}|}\hline
        \textit{\ubranch{} type} & \textit{micro-assembly} & \textit{prediction, first or second in triad} & \textit{prediction, last in triad} & \textit{confirmed microarchitectures} \\\hline
        cond. direct & \texttt{UJMPcc $\texttt{reg}_\texttt{cond}$,$\texttt{imm}_\texttt{tgt}$} & statically not-taken & statically not-taken & Goldmont, \textbf{Goldmont Plus, Haswell, Kaby Lake, Golden Cove} \\\hline
        uncond. direct & \texttt{UJMP $\texttt{imm}_\texttt{tgt}$} & \textbf{statically not-taken} & statically taken & Goldmont \\\hline
        cond. indirect & \texttt{UJMPcc $\texttt{reg}_\texttt{cond}$,$\texttt{reg}_\texttt{tgt}$} & \textbf{statically not-taken} & statically not-taken & Goldmont \\\hline
        uncond. direct & \texttt{UJMP $\texttt{reg}_\texttt{tgt}$} & \textbf{statically not-taken} & {\st{statically taken} \newline \textbf{not predicted}} & Goldmont \\\hline
    \end{tabular}
    \caption{Microsequencer's prediction of microcode branches based on position in triad, based on our experimentation.
    Plain text indicates reaffirmed behaviors previously observed in prior work~\cite{hackers-ucodedisasm}.
    \textbf{Bold text} indicates newly observed behaviors in this work.
    \st{Strikethrough text} represents refuted behaviors suggested in prior work~\cite{hackers-ucodedisasm}.}
    \label{tab:ubranch-semantics}
\end{table*}

\subsection{Microcode Branches}
Intel Goldmont microcode contains \textit{four types} of microcode branches, originally discovered by Ermolov et al.~\cite{hackers-writeup}:
unconditional direct branches (\texttt{UJMP $\texttt{imm}_\texttt{tgt}$}),
conditional direct branches (\texttt{UJMPcc $\texttt{reg}_\texttt{cond}$, $\texttt{imm}_\texttt{tgt}$}),
unconditional indirect branches (\texttt{UJMP $\texttt{reg}_\texttt{tgt}$}),
and conditional indirect branches (\texttt{UJMPcc $\texttt{reg}_\texttt{cond}, \texttt{reg}_\texttt{tgt}$}).

\paragraph{Sequential semantics}
Unconditional branches (i.e., \texttt{UJMP $\texttt{imm}_\texttt{tgt}$} and \texttt{UJMP $\texttt{reg}_\texttt{tgt}$}) always transfer control to the microcode address held in the register $\texttt{reg}_\texttt{tgt}$ or given in the immediate value $\texttt{imm}_\texttt{tgt}$.
Conditional branches (i.e., \texttt{UJMPcc $\texttt{reg}_\texttt{cond}$,$\texttt{imm}_\texttt{tgt}$} and \texttt{UJMPcc $\texttt{reg}_\texttt{cond}$,$\texttt{reg}_\texttt{tgt}$})
conditionally transfer control to the microcode address held in $\texttt{reg}_\texttt{tgt}$/$\texttt{imm}_\texttt{tgt}$
if and only if the condition register $\texttt{reg}_\texttt{cond}$'s flags satisfy the condition code \texttt{cc} (encoded in the micro-op's opcode).

\subsection{Prediction of Microcode Branches}
Ermolov et al.~\cite{hackers-ucodedisasm} first discovered that \textit{conditional direct/indirect \ubranches{} are statically predicted not-taken on the Goldmont microarchitecture}.
Motivated by this observation, we systematically evaluate the prediction of \textit{all four types of \ubranches{}} on Goldmont and other microarchitectures.
We summarize our novel findings in \Cref{tab:ubranch-semantics} and as follows:
\begin{itemize}
    \item Conditional direct \ubranches{} are statically predicted not-taken on all Intel microarchitectures we evaluate (Goldmont, Goldmont Plus, Haswell, and Golden Cove; \S\ref{sec:ubranches:pred:cond}).
    \item All \ubranches{}, including unconditional ones, are statically predicted not-taken on Goldmont if they are not the last in their triad (\S\ref{sec:ubranches:pred:uncond-dir}).
    \item Unconditional indirect \ubranches{} are not predicted if they are the last of their triad, contrary to Ermolov et al.'s suggestion~\cite{hackers-ucodedisasm} (\S\ref{sec:ubranches:pred:uncond-ind}).
   
\end{itemize}

\subsubsection{Prediction of conditional microcode branches}\label{sec:ubranches:pred:cond}
Ermolov et al.~\cite{hackers-ucodedisasm} observe that \textit{conditional} direct and indirect microcode branches are \textit{always statically predicted not-taken under all circumstances} on Goldmont cores.
We have run many experiments on our Goldmont core that reaffirm this.
Moreover, based on our experiments, this behavior appears consistent across modern Intel processors: 
all cores we evaluated exhibit \textit{static prediction of conditional microcode branches}.

Below, we describe one example of how we have inferred the existence of a statically-predicted conditional microcode branch in the \texttt{cld} instruction for Intel microachitectures other than Goldmont, with no public microcode available.

\paragraph{Example: \texttt{cld}.}
The \texttt{cld} (``Clear Direction Flag'') instruction, clears the direction (\texttt{DF}) flag in x86's RFLAGS register~\cite{intel-sdm}.
The direction flag \texttt{DF} determines the direction of x86 string operations, like \texttt{movsb}, which performs a byte-wise copy of memory from \texttt{rsi} to \texttt{rdi}. 
If \texttt{DF} is reset (to $0$), then \texttt{rsi} and \texttt{rdi} are incremented after each byte copy;
if \texttt{DF} is set (to $1$), then \texttt{rsi} and \texttt{rdi} are decremented.
Using the \verb|UOPS_RETIRED.MS| performance counter, we confirm that \texttt{cld} is microcoded on all Intel cores we evaluate. 

\subparagraph{\indent \rm \it Hypothesis:}
We expect to find a conditional \ubranch{} in \texttt{cld}'s microprogram for the following reasons.
First, \texttt{DF} is a non-arithmetic flag, which is unlikely to be renamed and thus likely requires full serialization of the instruction stream to modify it (e.g., \cref{line:cld-pseudo:write-df} in \Cref{lst:cld-pseudo} does not execute until all prior instructions have executed, and no subsequent instructions execute until \cref{line:cld-pseudo:write-df} has executed).
Thus, if \texttt{DF} is set, then \texttt{cld} must write zero to \texttt{DF} and serialize execution (slow path).
However, if \texttt{DF} is already cleared, then \texttt{cld} need not do anything (fast path).
Thus, we hypothesize that all Intel microarchitectures implement \texttt{cld} using a microcode branch conditioned on \texttt{DF}. 

\subparagraph{\indent \rm \it Goldmont microprogram:}
To start, using an adaptation of the microcode tracing technique proposed by Borrello et al.~\cite{custom-processing-unit}, we successfully identified the microprogram implementing \texttt{cld} on Goldmont, reproduced in \Cref{lst:cld}, which indeed contains a conditional \ubranch{} (\textbf{bolded}).
\Cref{lst:cld-pseudo} presents higher-level pseudocode representing the microprogram in \Cref{lst:cld}.
Note that \verb|unk_abc| indicates a micro-op with an unknown opcode \verb|0xabc|. 

\begin{figure}
\begin{subfigure}{\linewidth}
\begin{lstlisting}[style=example, escapeinside={?}{?}, caption={Goldmont microprogram implementing \texttt{cld}}, label=lst:cld]
cld:
  tmp1 = unk_109(1)?\label{line:cld:pred}? // reads DF into tmp1
  ?\textbf{UJMPC(tmp1, .slow)}?
.fast:
  NOP
  SEQW UEND0

.slow:
  tmp5 = RDCREG64(RFLAGS)
  BTS_WRCREG64(tmp5, 10, RFLAGS)?\label{line:cld:update}?
  NOP
  SEQW SYNCFULL
  unk_256(0)
  SEQW LFNCEWAIT
  SEQW UEND0
  
\end{lstlisting}
\caption{Simplified microprogram}
\label{fig:cld:ucode}
\end{subfigure}

\vspace{5mm}

\begin{subfigure}{\linewidth}
\begin{lstlisting}[style=example, caption={Pseudocode for \texttt{cld}}, escapeinside={?}{?}, label=lst:cld-pseudo]
if (DF) {
  RFLAGS.DF = 1;?\label{line:cld-pseudo:write-df}?
  serialize();
}
\end{lstlisting}
\caption{Microprogram pseudocode}
\label{fig:cld:pseudo}
\end{subfigure}
\caption{Microprogram implementing \texttt{cld} on Goldmont, located at MSROM address \texttt{U06d0} in Goldmont's microcode disassembly~\cite{hackers-ucodedisasm}. We use CustomProcessingUnit~\cite{custom-processing-unit} to verify this mapping.}
\label{fig:cld}
\end{figure}

\subparagraph{\indent \rm \it Inferring prediction semantics via nanobenchmarks:}
To confirm that \texttt{cld}'s conditional \ubranch{} in \Cref{lst:cld} is statically predicted not-taken, 
we craft targeted nanobenchmarks and run them under Abel et al.'s nanoBench tool~\cite{nanobench} on a Goldmont core. %
Given a snippet of assembly or machine code---i.e., the \textit{nanobenchmark}---and a requested performance counter event, nanoBench replicates the snippet many times (1000 times by default), executes the replicated code in a loop, and reports the average number of events per executed snippet. %
We run the nanobenchmarks described below and monitor the \verb|UOPS_ISSUED.ANY| and \verb|UOPS_RETIRED.ANY| performance counters~\cite{intel-sdm} to determine
how many micro-ops are issued and retired, respectively, per snippet.

\textit{Nanobenchmark 1.}
To start, we run the simple nanobenchmark consisting of the single instruction (macro-op) ``\texttt{cld}'' (expanded to ``\texttt{for (i=0;i<n;++i)\{cld;cld;cld;\ldots\}}'' by nanoBench) on our Goldmont core.
We expect \texttt{cld}'s microprogram to take the fast path every time, since \texttt{DF} is clear by default, and thus the branch should not be taken.
As expected, we observe that exactly three micro-ops issue and retire per iteration, corresponding to the fast path in \Cref{lst:cld}.
This indicates that no micro-ops along the slow path were transiently issued, thus the branch was conclusively not mispredicted taken.
We also observe three micro-ops issued and retired on all other Intel cores we evaluate.

\textit{Nanobenchmarks 2 and 3.}
However, the above results do not prove that the branch was predicted not-taken, let alone that it is \textit{statically} (i.e., always) predicted not-taken.
To prove this, we run nanobenchmark 2 ``\texttt{std; lfence}'' and nanobenchmark 3 ``\texttt{std; lfence; cld}.'' 
By comparing the difference in issued/retired micro-ops between these two nanobenchmarks, we can determine if introducing \texttt{cld} in nanobenchmark 3 introduces a \ubranch{} misprediction.
\texttt{std} is the ``set direction flag'' instruction, and we use \texttt{lfence} to serialize the instruction stream.
For nanobenchmark 2, we observe 7/7 micro-ops issued/retired;
for nanobenchmark 3, we observe 17/13 micro-ops issued/retired.
Thus, \texttt{cld} in nanobenchmark 3 causes the issue/retirement of +10/+6 additional micro-ops.
Four issued micro-ops are not retired and thus are \textit{transiently issued}, implying a microcode branch misprediction of \texttt{cld}.
This proves that \texttt{cld}'s microcode branch is \textit{always} transiently mispredicted not-taken, 
causing the transient issue of the two remaining micro-ops along the fast path
and the first two micro-ops of the next macro-op (\texttt{std}).

\subparagraph{\rm \it Conclusions:}
We ran the aforementioned nanobenchmarks on all three other Intel cores we consider (\S\ref{sec:ubranches:cores}),
and we observed similar results.
Thus, we draw the following conclusions.

First, on all evaluated Intel cores, \texttt{cld} is microcoded using a \textit{conditional \ubranch{}} that introduces a fast path of exactly three micro-ops and a slow path of a platform-dependent number of micro-ops.

Second, all studied cores \textit{statically predict} the outcome of said conditional \ubranch{}.
On Goldmont, we are certain \texttt{cld}'s \ubranch{} is statically predicted not-taken.
We cannot be certain that this is also the case for the other cores whose microcode is not public---hypothetically, all \ubranches{} could be statically predicted taken instead.
However, we see no reason why Intel would make this opposite design choice for some cores but not others.
Thus, we make the following claim with high confidence:
\begin{tcolorbox}[colback=blue!5!white,colframe=blue!75!black,title=Observation 1, label=sec:ubranches:obs1]
    Intel cores statically predict that conditional microcode branches are not-taken.
\end{tcolorbox}

\subsubsection{Prediction of Unconditional Direct Microcode Branches}
\label{sec:ubranches:pred:uncond-dir}
We are the first to observe that on Goldmont cores, \textit{unconditional direct} \ubranches{} are statically (mis)pre\-dicted not-taken if they are not the last micro-op in their triad (\Cref{tab:ubranch-semantics}).
(If they are the last in their triad, they are always predicted taken, as previously observed by Ermolov et al.~\cite{hackers-ucodedisasm}.)
We are unsure if this behavior is also present on other Intel cores.
In the Goldmont microcode, we found multiple unconditional branches that are not the last in their triad.
However, all of these were directly followed by the \texttt{SEQW SYNCFULL} microsequencer directive (\S\ref{sec:background:seqw}), which instructs the microsequencer to stop fetching micro-ops until all prior micro-ops have retired, thus acting as a speculation fence.
Instead, we verified this behavior by introducing vulnerable unconditional direct \ubranches{} via microcode patches.

\subsubsection{Unconditional Indirect Microcode Branches}\label{sec:ubranches:pred:uncond-ind}
On Goldmont cores (and likely other Intel cores), the targets of unconditional indirect microcode branches that are the last in their triad are \textit{never predicted}.
Furthermore, they do not execute at least until all prior %
conditional (and possibly unconditional) \ubranches{} have resolved (and possibly not until retirement). 
We suspect that indirect microcode branches wait for prior \ubranches{} to resolve for safety reasons. 
For example, a mispredicted \ubranch{} may cause a subsequent indirect \ubranch{} to jump to the wrong target and transiently execute
unbuffered writes to microarchitectural state that persist even if the writes are subsequently squashed.
Like Ermolov et al.~\cite{hackers-writeup}, we observe multiple types of micro-ops that perform such unbuffered writes, including control register writes (\texttt{WRCREG}) and micro-RAM writes (\texttt{WRURAM}).

\subsubsection{Detection and Resolution of Microcode Branch Predictions}
On all Intel cores we have studied, we observe that while they are detected out-of-order,
\textit{all \ubranch{} mispredictions are resolved (i.e., corrected) in-order}.
In fact, all of our observations support the conclusion that \textit{mispredictions are resolved when the \ubranch{} retires}.
However, when the microsequencer detects a \ubranch{} misprediction, it appears to stop issuing micro-ops along the mispredicted path to the execution backend.
This narrows the transient execution window following \ubranch{} mispredictions whose predicates resolve quickly (i.e., depend on immediately available data).
Even so, as we discuss in the following section (\S\ref{sec:ubranches:window}), these transient execution windows are still large enough to leak sensitive data.

\subsection{Transient Execution Following Microcode Branch Mispredictions}\label{sec:ubranches:window}
Not only are conditional microcode branches statically predicted not-taken, but 
we also observe that transient execution windows following mispredicted branches
can span several subsequent macro-ops.

\begin{figure*}
\begin{subfigure}[b]{0.45\textwidth}
\begin{lstlisting}[style=example, escapeinside=??, numbers=left]
?\lstseq{meb\_vulnerable\_inst:}?
  ?\lstseq{pred = f(input arch state) // resolves to true}?
  ?\lstbr{UJMPcc(pred, .exception) // check for exception}??\label{line:meb-vuln:br}?
  ?\lsttrans{...}?
  ?\lsttrans{[\hl{data} = access\_secret()]}??\label{line:meb-vuln:access}?
  ?\lsttrans{[\hl{out\_arch\_reg} = \hl{data}]}??\label{line:meb-vuln:expose}?
  ?\lsttrans{SEQW UEND0}?

.exception:?\label{line:meb-vuln:except}?
  SIGEVENT(event)
  ...
  SEQW UEND2
\end{lstlisting}
\caption{\meb{} vulnerability pattern (microcode)}
\label{fig:meb:vuln}
\end{subfigure}\hfill
\begin{subfigure}[b]{0.45\textwidth}
\begin{lstlisting}[style=example, escapeinside=??, numbers=left]
void attack1() {?\label{line:meb-attack1-begin}?
  ?\lstseq{u64 data;}?
  ?\lstseq{// sequentially faults,}?
  ?\lstseq{// but transiently returns secret}?
  ?\lstbr{asm ("meb\_vulnerable\_inst \%0" : "=r"(\hl{data}));}?
  ?\lsttrans{leak(\hl{data});}??\label{line:meb-attack1-leak}\label{line:meb-attack1-end}?
}
void attack2() {?\label{line:meb-attack2-begin}?
  ?\lstseq{// sequentially faults, but transiently}?
  ?\lstseq{// allows following macro-ops to access secret}?
  ?\lstbr{asm ("meb\_vulnerable\_inst \%0" :: "r"(idx);}??\label{line:meb-attack2:vuln}?
  ?\lsttrans{u64 *addr = compute\_address(idx);}??\label{line:meb-attack2:idx2}?
  ?\lsttrans{u64 \hl{data} = *addr;}? // returns secret
  ?\lsttrans{leak(\hl{data});}??\label{line:meb-attack2-leak}\label{line:meb-attack2-end}? // transmitter
}
\end{lstlisting}
\caption{\meb{} attack templates (architectural programs)}
\label{fig:meb:attack}
\end{subfigure}
\caption{Structure of \meb{} (a) vulnerabilities and (b) attacks.}
\end{figure*}

Specifically, we observe that all evaluated Intel cores transiently execute \textit{up to 12 macro-ops} following the mispredicted conditional \ubranch{} in \verb|cld|, even though its predicate only depends on one micro-op (\cref{line:cld:pred}) which executes in a single cycle.
One some cores, this transient macro-op execution window is much larger: for example, we observed that Goldmont Plus may transiently execute up to 22 macro-ops following the conditional \ubranch{} misprediction in \texttt{cld}. %

Furthermore, if a conditional \ubranch{} predicate depends on long-latency macro-op operands, we observe a macro-op transient execution window that appears to be limited only by the size of the reorder buffer.
From these results, we make the following observation:
\begin{tcolorbox}[colback=blue!5!white,colframe=blue!75!black,title=Observation 2, label=sec:ubranches:obs2]
    All evaluated Intel cores can transiently execute several macro-ops following a conditional \ubranch{} misprediction, regardless of the latency of its predicate.
    If the predicate is long-latency, they can execute as many subsequent macro-ops as fit in the ROB.
\end{tcolorbox}
\noindent Now, we turn to exploiting these transient execution windows following \ubranch{} mispredictions in \S\ref{sec:uspectre}.

\section{\Ourspectre{} Attacks}
\label{sec:uspectre}

\begin{figure*}[t]
\begin{subfigure}[b]{0.45\textwidth}
\begin{lstlisting}[style=example, escapeinside=??]
mvi_vulnerable_inst:
  ?\lstseq{pred = f(input arch state) // resolves to true}??\label{line:mvi-vuln-pred}?
  ?\lstbr{UJMPcc(pred, .slow) // check for slow path}??\label{line:mvi-vuln-branch}?
?\lsttrans{.fast:}??\label{line:mvi-vuln-fast-begin}?
  ?\lsttrans{[output arch state = incorrect result]}??\label{line:mvi-vuln-update}?
  ?\lsttrans{SEQW UEND0}??\label{line:mvi-vuln-fast-end}?
.slow:
  output arch state = correct result
  SEQW UEND0  
\end{lstlisting}
\caption{\mvi{} vulnerability pattern (microcode)}
\label{fig:mvi:vuln}
\end{subfigure}\hfill
\begin{subfigure}[b]{0.45\textwidth}
\begin{lstlisting}[style=example, escapeinside=??]
void mvi_attack() {
  ?\lstseq{u64 input = ...;}??\label{line:mvi-attack:prime}?
  ?\lstseq{u64 output;}?
  ?\lstbr{asm ("mvi\_vulnerable\_inst \%0"}?
       ?\lstbr{: "=r"(output) : "r"(input));}?
  ?\lsttrans{u64 *addr = compute\_address(output);}??\label{line:mvi-attack:tbegin}\label{line:mvi-attack:addr}?
  ?\lsttrans{u64 \hl{data} = *addr;}??\label{line:mvi-attack:access}? // secret access
  ?\lsttrans{leak(\hl{data});}??\label{line:mvi-attack:tend}\label{line:mvi-attack:leak}? // transmitter
}
\end{lstlisting}
\caption{\mvi{} attack pattern (architectural program)}
\label{fig:mvi:attack}
\end{subfigure}
\caption{Structure of \mvi{} (a) vulnerabilities and (b) attacks.}
\label{fig:mvi}
\end{figure*}

We present \textit{\ourspectre{}}, a new class of transient execution attacks that exploit transient execution following conditional \ubranch{} mispredictions to leak sensitive or unauthorized data.
Many existing Spectre and Meltdown attacks are actually instances of \ourspectre{} attacks on some or all the Intel microarchitectures we study; we will discuss these attacks in \S\ref{sec:existing}.
In this section, we define \uspectre{} attacks (\S\ref{sec:uspectre:uspectre}), describe three variants of \uspectre{} attacks (\S\ref{sec:uspectre:meb}--\ref{sec:uspectre:mil}),
and incorporate \uspectre{} into the established transient execution attack taxonomy (\S\ref{sec:uspectre:taxonomy}).

In this paper, we focus on exploiting conditional \textit{direct} \ubranch{} mispredictions.
However, \uspectre{} can also exploit conditional \textit{indirect} branch mispredictions (which comprise 6.0\% of conditional branches in Goldmont's microcode disassembly~\cite{hackers-ucodedisasm}).
\ourspectre{} cannot exploit \textit{unconditional} indirect branches, as they are not predicted (i.e., are executed at retirement).

\subsection{\Ourspectre{} Vulnerabilities}
\label{sec:uspectre:uspectre}
An architectural instruction's (macro-op's) microprogram is vulnerable to \ourspectre{} if it contains a conditional \ubranch{}
whose misprediction can be exploited by an architectural program to leak sensitive architectural or microarchitectural state without authorization. %
Some types of \ourspectre{} vulnerabilities, such as \mvi{}-ZDI (\S\ref{sec:existing:zdi}) or \meb{}-FSGSBASE (\S\ref{sec:existing:fsgsbase}) masquerade as apparent instances of Spectre or Meltdown, depending on the consequences of the vulnerable \ubranch{} misprediction.
Other \uspectre{} vulnerabilities do not resemble any existing transient execution attacks at all, like \uspectre{}-MEB-PCIDX (\S\ref{sec:new:pcidx}).

\subsection{\uspectre{}-MEB: Microcode Exception Bypass}
\label{sec:uspectre:meb}
The first variant of \ourspectre{} attacks we identify is \textit{microcode exception bypass}, or \textit{\uspectre{}-MEB} for short.
These attacks exploit a conditional \ubranch{} whose \textit{taken} path raises a hardware exception.
Such \ubranches{} are vulnerable because even if the exception condition is satisfied, the microsequencer statically mispredicts the \ubranch{} \textit{not-taken} (per Observation 1, \S\ref{sec:ubranches:obs1}).
This creates a transient execution window that may span into subsequent macro-ops (per Observation 2, \S\ref{sec:ubranches:obs2}), allowing an attacker to leak sensitive data that was meant to be protected by raising the exception.

\subsubsection{MEB Vulnerability and Attack Details}
\label{sec:uspectre:annotations}
\Cref{fig:meb:vuln} shows the structure of microprograms that are vulnerable to \meb{}.
The predicted execution path through the microprogram is \textbf{bold};
sequentially executed instructions are plain text (black and not bolded);
the vulnerable, mispredicted \ubranch{} is \textcolor{red}{red};
and transiently executed instructions following the mispredicted \ubranch{} are \textcolor{gray}{gray}.
[Bracketed] lines are not present in all MEB vulnerabilities.
\hl{Highlighted expressions} contain sensitive data.
We use these conventions throughout the rest of the paper.

The vulnerable \ubranch{} on \cref{line:meb-vuln:br} sequentially branches to an exception routine (\verb|.exception|, \cref{line:meb-vuln:except}) but is mispredicted not-taken (\S\ref{sec:ubranches:pred:cond}) and transiently executes the rest of the microprogram and subsequent macro-ops.
Some MEB vulnerabilities (e.g., \meb{}-REG, \S\ref{sec:existing:rsrr}) but not others (e.g., \meb{}-OF, \S\ref{sec:new:of}) access sensitive data (\cref{line:meb-vuln:access}) and expose it via an output architectural register of the microprogram (\cref{line:meb-vuln:expose}) following the mispredicted \ubranch{}.
(We will discuss how this can be exploited to leak sensitive data next.)

\Cref{fig:meb:attack} shows two attack patterns that can be used to exploit \meb{} vulnerabilities.
Attack pattern 1 (\texttt{attack1}, \crefrange{line:meb-attack1-begin}{line:meb-attack1-end}) can be used to exploit MEB-vulnerable microprograms that transiently expose sensitive data
to subsequent macro-ops via an architectural output of the vulnerable macro-op (\cref{line:meb-vuln:access,line:meb-vuln:expose} in \Cref{fig:meb:vuln}).
Attack pattern 2 (\texttt{attack2}, \crefrange{line:meb-attack2-begin}{line:meb-attack2-end} in \Cref{fig:meb:attack}) can be used to exploit other MEB vulnerabilities,
which allow subsequent macro-ops to both transiently access and leak sensitive architectural state (e.g., secret memory locations) that they otherwise would not.
The variable \verb|idx| represents any invalid data that (a) triggers the exception in the MEB-vulnerable macro-op and (b) is used to transiently access out-of-bounds secret data that leaks.
In both cases, the attacker can leak secret data via a transmitter (\S\ref{sec:background:hw-side-channels}).

\subsubsection{Comparison to Meltdown}
From the perspective of the architectural program, \uspectre{}-MEB attacks appear misleadingly to exploit Meltdown-style delayed fault handling.
Thus, we say that \uspectre{}-MEB is \textit{Meltdown-presenting}.
However, there are key differences between \uspectre{}-MEB and Meltdown attacks.
In Meltdown attacks (\S\ref{sec:background:meltdown}), a faulting micro-op partially executes, but delayed handling of the fault allows subsequent instructions to transiently execute.
In contrast, no faulting instruction enters the instruction stream in the first place during a \meb{} attack, since the exception path was predicted not-taken.
One notable similarity is that in both Meltdown attacks and microcode exception bypass attacks, 
macro-ops following the faulting macro-op \textit{transiently execute} but do not architecturally commit.

\subsection{\mvi{}: Microcode Value Injection}
\label{sec:uspectre:mvi}
The second variant of \ourspectre{} attacks we identify is \textit{microcode value injection}, or \textit{\uspectre{}-MVI} for short.
These attacks involve a mispredicted \ubranch{} that causes architecturally incorrect results to be transiently exposed to subsequent macro-ops, i.e., it transiently \textit{injects} ``bad'' values into the architectural program's subsequent computation.
This allows subsequent macro-ops to transiently access and leak sensitive data that they otherwise would not.
Unlike in \meb{}, the vulnerable macro-op containing the mispredicted \ubranch{} does not architecturally fault. %
Instead, once the \ubranch{} misprediction is resolved, execution resumes along the correct, taken path of the \ubranch{}, and any transiently executed macro-ops are re-executed with the correct outputs of the microprogram.

\Cref{fig:mvi:vuln} shows the structure of a \mvi{} vulnerability.
Based on the current architectural state, the vulnerable microprogram computes (\cref{line:mvi-vuln-pred}) whether to take the fast path or slow path, which correspond to the not-taken and taken directions of the vulnerable \ubranch{} (\cref{line:mvi-vuln-branch}), respectively.
Because the \ubranch{} is always predicted not-taken (Observation 1, \S\ref{sec:ubranches:obs1}), 
the microprogram will transiently execute the fast path (\crefrange{line:mvi-vuln-fast-begin}{line:mvi-vuln-fast-end}).
This transient path either updates architectural state with data that is invalid due to the misprediction (\cref{line:mvi-vuln-update}) or, in some cases, does not update architectural state when it ought to (as in \texttt{cld}'s microprogram, \Cref{lst:cld}, which only updates \texttt{DF} on the slow path in \cref{line:cld:update} but not on the fast path).
After the end of the microprogram's fast path is reached (\cref{line:mvi-vuln-fast-end}), 
the processor transiently executes subsequent macro-ops (Observation 2, \S\ref{sec:ubranches:obs2}),
which leverage the invalid architectural state to transiently access and leak sensitive data.

\mvi{} can be exploited to leak sensitive data via the attack pattern shown in \Cref{fig:mvi:attack}.
First, the attacker primes the input architectural state in order to trigger the misprediction (\cref{line:mvi-attack:prime}).
Then, the program executes an instruction whose microprogram is vulnerable to \mvi{},
causing the transient execution of subsequent macro-ops on invalid results.
These subsequent macro-ops use these invalid results to compute (\cref{line:mvi-attack:addr}) and dereference (\cref{line:mvi-attack:access}) a pointer to sensitive data, which is then leaked (\cref{line:mvi-attack:leak}).

Note that the transiently executed instructions following the vulnerable instruction will be re-executed using the correct results (produced by the microprogram's slow path) when the \ubranch{} misprediction resolves (unlike in \meb{}).

\paragraph{Comparison to Spectre}
From the perspective of the architectural program, \mvi{} resembles Spectre-style mispredictions, specifically value mispredictions, hence its name;
MVI also has a similar leakage potential to Spectre. 
Hence, we say that \mvi{} attacks are \textit{Spectre-presenting}.
However, microcode value injections are not value mispredictions from a microarchitectural perspective, which involve predicting the output of a single micro-op, like a load~\cite{pandora}.

\begin{figure*}[t]
\begin{subfigure}[b]{0.45\textwidth}
\begin{lstlisting}[style=ucode, escapeinside=??]
mil_vulnerable_inst:
  ?\lstseq{pred = ... // resolves to true}?
  ?\lstbr{UJMPcc(pred, .noleak)}??\label{line:mil-vuln:br}?
?\lsttrans{.leak:}?
  ?\lsttrans{\hl{data} = access(state)}??\label{line:mil-vuln:access}? // secret access
  ?\lsttrans{leak(\hl{data})}??\label{line:mil-vuln:leak}? // transmitter
  ...
.noleak:
  ...
\end{lstlisting}
\caption{\mil{} vulnerability pattern (microcode)}
\label{fig:mil:vuln}
\end{subfigure}\hfill
\begin{subfigure}[b]{0.45\textwidth}
\begin{lstlisting}[style=mcode, escapeinside=??]
void mil_attack() {
  ?\lstseq{[prime\_microarchitectural\_state();]}??\label{line:mil-attack:prime}?
  ?\lstbr{mil\_vulnerable\_inst(/*inputs*/...);}??\label{line:mil-attack:inst}?
}
\end{lstlisting}
\caption{\mil{} attack pattern (architectural program)}
\label{fig:mil:attack}
\end{subfigure}
\caption{Structure of a \mil{} (a) vulnerabilities and (b) attacks.}
\label{fig:mil}
\end{figure*}

\subsection{\mil{}: Microcode-Internal Leaks}
\label{sec:uspectre:mil}
The final \uspectre{} variant we consider is \textit{microcode-internal leaks}, or \textit{\mil{}} for short, in which a vulnerable microprogram both transiently accesses (\cref{line:mil-vuln:access} in \Cref{fig:mil:vuln}) and leaks (\cref{line:mil-vuln:leak}) sensitive architectural or microarchitectural state due to a prior \ubranch{} misprediction (\cref{line:mil-vuln:br}). %

\mil{} vulnerabilities can be exploited using the simple attack pattern shown below in \Cref{fig:mil:attack}.
First, the attacker may need to prime microarchitectural state like the cache (\cref{line:mil-attack:prime}), as in \mil{}-SCO (\S\ref{sec:existing:sco}).
Then, the attacker invokes the MIL-vulnerable instruction with carefully selected inputs that will induce the microprogram to transiently access and leak secret data (\cref{line:mil-attack:inst}) via a transient transmitter following the vulnerable instruction.

\paragraph{\mil{} Transmitters}
In some sense, microcode-internal leaks transform the macro-op they implement into a transmitter (\S\ref{sec:background:transmitter}).
What makes MIL transmitters more severe than traditional transmitters, however, is their ability to \textit{transiently access and leak} state that they should not access architecturally. %
In contrast, traditional transmitters can only, at worst, leak their architectural operands. %

\paragraph{Mitigating MIL}
Unlike in MEB and MVI, which require subsequent macro-ops to transiently leak secrets, 
\mil{} attacks cause secret data to be both accessed \textit{and} leaked in the same microprogram.
\textit{As a result, \mil{} attacks cannot be blocked by inserting a speculation fence after the vulnerable instruction.}
Instead, privacy-sensitive programs in many cases must eschew using the vulnerable instruction altogether to avoid leaking private data.

In \S\ref{sec:uspectre:mil}, we provide an example of a microcode-internal leak present on Intel Goldmont cores (and likely all other Intel cores).

\subsection{Extending the Transient Execution Attack Taxonomy}
\label{sec:uspectre:taxonomy}

\begin{figure}
    \centering
    \includegraphics[width=\linewidth]{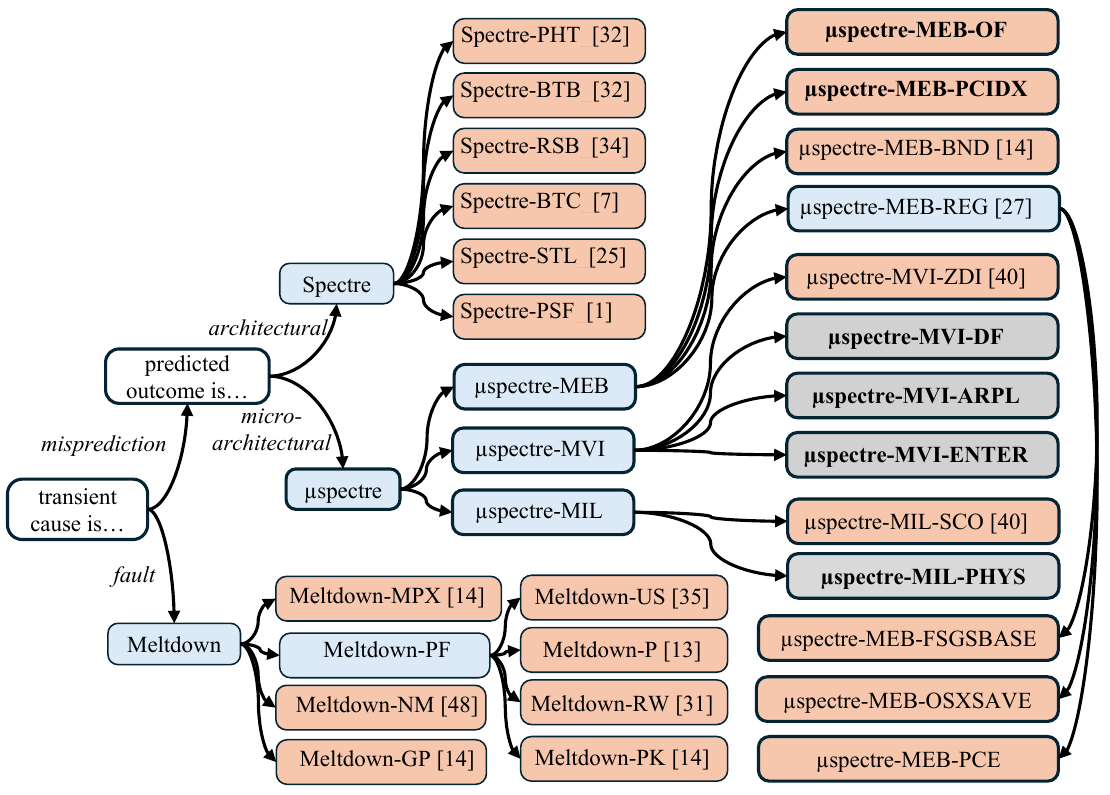}
    \caption{Transient execution attack classification tree, simplified from prior work \cite{canella-systematic} and adapted to include \uspectre{} and other Spectre variants that have been since discovered~\cite{spectre-btc, phantom, spectre-psf}.
    Demonstrated attacks are in red.
    Presently inconclusive results are in gray.
    \textbf{Bolded text} represents a new vulnerability presented in this paper.
    }
    \label{fig:taxonomy}
\end{figure}

We extend the transient execution attack taxonomy presented by Canella et al.~\cite{canella-systematic} 
to incorporate \uspectre{} attacks
as follows.
\Cref{fig:taxonomy} visualizes this taxonomy and presents all \uspectre{} attacks/vulnerabilities considered in this work.

\paragraph{Meltdown attacks}
Meltdown attacks~\cite{meltdown, foreshadow, lvi, mds-fallout, mds-ridl, mds-zombieload, lazyfp}
exploit transient execution following a \textit{faulting micro-op}.
Meltdown attacks typically exploit \textit{delayed handling} of the fault, which permits subsequent macro-ops to transiently access and/or leak sensitive data before the CPU squashes micro-ops following the fault~\cite{canella-systematic}.

\paragraph{Prediction-based transient execution attacks}
All other transient execution attacks exploit various types of mispredictions to transiently access and/or leak sensitive data before the misprediction is resolved.
Canella et al.~\cite{canella-systematic} consider all prediction-based transient execution attacks to be Spectre.
However, \uspectre{} is fundamentally different from Spectre; for example, it can masquerade as a Meltdown variant.
Thus, we refine prediction-based transient execution attacks into two subclasses based on whether the \textit{state being predicted} is \textit{architectural} or \textit{microarchitectural}. %

\subparagraph{\indent Spectre: architectural state predictions.} \textit{Spectre attacks} exploit \textit{(mis)predictions of architectural state} following an instruction or micro-op. 
For example, branch predictions---exploited by Spectre-PHT, Spectre-BTB~\cite{spectre}, Spectre-RSB~\cite{spectre-rsb}, and Spectre-BTC~\cite{phantom, spectre-btc}---predict the architectural program counter following a macro-op.
Spectre-STL and Spectre-PSF predict the architectural data returned by a load.

\subparagraph{\indent \uspectre{}: microarchitectural state predictions.} In contrast, \textit{\uspectre{} attacks} exploit \textit{(mis)predictions of microarchitectural state}.
In this paper, we focus exclusively on mispredictions of the microprogram counter (\S\ref{sec:background:upc}).
We leave it to future work to explore whether other types of microarchitectural state can be predicted.

\section{Reclassifying Existing Attacks as \Ourspectre{}}
\label{sec:existing}

Many existing transient execution attacks previously considered to be Meltdown or Spectre variants are, in fact, instances of \ourspectre{}.
Specifically, we find that Rogue System Register Read~\cite{intel-rogue-system-register-read} and Meltdown-BND~\cite{canella-systematic}, originally believed to be Meltdown variants, are actually instances of \meb{} vulnerabilities, on some Intel processors.
Thus, we refer to them as \meb{}-REG (\S\ref{sec:existing:rsrr}) and \meb{}-BND (\S\ref{sec:existing:bnd}), respectively.
We also find that zero dividend injection (ZDI)~\cite{revizor++} is an instance of \mvi{} and string comparison overrun (SCO)~\cite{revizor++} is an instance of \mil{}.
We refer to these as \mvi{}-ZDI (\S\ref{sec:existing:zdi}) and \mil{}-SCO (\S\ref{sec:existing:sco}), respectively.

\subsection{Rogue System Register Read (\meb{}-REG)}
\label{sec:existing:rsrr}
We are the first to report that some Intel microarchitectures, many \textit{Rogue System Register Read}~\cite{intel-rogue-system-register-read} attacks are in fact instances of \textit{microcode exception bypass} (\meb{}, \S\ref{sec:uspectre:meb}), not Meltdown, contrary to the suggestion of prior work~\cite{canella-systematic}.
We call the class of \meb{} vulnerabilities that enable Rogue System Register Read \textit{\meb{}-REG}.

\paragraph{Background}
In a Rogue System Register Read attack, an unprivileged attacker executes a vulnerable instruction that transiently bypasses a privilege check to read privileged state (e.g., a system register) and pass it to a subsequent transient transmitter, leaking it.

In the past, this transient behavior has been categorized as a Meltdown-style attack due to delayed fault handling~\cite{canella-systematic} (\S\ref{sec:uspectre:taxonomy}).
However, we have discovered this is not case on some Intel processors; rather, we observe this behavior is due to a \textit{microcode exception bypass}.
To demonstrate this, we perform a case study of Rogue System Register Read vulnerabilities on the Goldmont microarchitecture, whose microcode is public~\cite{hackers-ucodedisasm}.

\subsubsection{Case Study: Intel Goldmont}
\label{sec:existing:rsrr:goldmont}
On Goldmont, we find that three instructions are vulnerable to Rogue System Register Read, i.e., \meb{}-REG: \texttt{rdfsbase}/\texttt{rdgsbase}, \texttt{xgetbv}, and \texttt{rdpmc}.
By examining the microprograms implementing them, %
we find that three of these implement privilege checks using a \textit{vulnerable conditional \ubranch{}} to an exception routine, %
constituting a textbook \uspectre{} vulnerability.
The misprediction of this \ubranch{} creates a transient execution window that extends into the following macro-ops in which an attacker can leak the privileged state.

Furthermore, we find that each vulnerable \ubranch{} is predicated on a control register bit that guards whether the microprogram should raise an exception.
Thus, we label subvariants of \meb{}-REG according to the name of the guarding control register bit.

\begin{figure}
\begin{subfigure}{\linewidth}
\begin{lstlisting}[style=example, escapeinside={?}{?}]
rdfsbase/rdgsbase r64:
  ?\lstseq{tmp0 = RDCREG64(CR4)}?
  ?\lstbr{BT\_UJMPNC(tmp0, 16, \#GP)}?
  ?\lsttrans{\lstsec{tmp2} = RDSEG(FS/GS, BASE)}?  // unauthorized access
  ?\lsttrans{\lstsec{r64} = ZEROEXTN(\lstsec{tmp2})}? // exposed via arch. output
  ?\lsttrans{SEQW UEND0}?
    
#GP:
  SIGEVENT(0x19)
  ...
\end{lstlisting}
\caption{MEB-vulnerable microprogram (simplified)}
\label{fig:rdfsbase:glm:ucode}
\end{subfigure}

\vspace{5mm}

\begin{subfigure}{\linewidth}
\begin{lstlisting}[style=example, escapeinside=??]
?\lstbr{if (!CR4.FSGSBASE)}?
  raise #GP;
?\lsttrans{\hl{r64} = \hl{FS\_BASE/GS\_BASE};}?
\end{lstlisting}
\caption{Pseudocode for vulnerable microprogram}
\label{fig:rdfsbase:glm:pseudo}
\end{subfigure}

\vspace{5mm}

\begin{subfigure}{\linewidth}
\begin{lstlisting}[style=example, escapeinside=??]
// assume CR4.FSGSBASE=0
?\lstseq{u64 fsbase;}?
?\lstbr{asm ("rdfsbase \%0" : "=r"(\hl{fsbase}));}?
?\lsttrans{transmit(\hl{fsbase});}?
\end{lstlisting}
\caption{Rogue System Register Read attack (architectural program)}
\label{fig:rdfsbase:glm:attack}
\end{subfigure}
\caption{MEB-vulnerable microprogram implementing \texttt{rdfsbase/rdgsbase} on Goldmont.}
\label{fig:rdfsbase:glm}
\end{figure}

\paragraph{\meb{}-FSGSBASE (\texttt{rdfsbase}/\texttt{rdgsbase})}
\label{sec:existing:fsgsbase}
It has been well known since 2018~\cite{intel-rogue-system-register-read} that the \texttt{rdfsbase} and \texttt{rdgsbase} instructions allow transient access to the \texttt{FS}/\texttt{GS} base registers, even if the instructions have been disabled by setting \texttt{CR4.FSGSBASE=0}~\cite{intel-sdm}.

Below, we include the vulnerable Goldmont microprogram implementing \texttt{rdfsbase}/\texttt{rdgsbase} (\Cref{fig:rdfsbase:glm:ucode}) along with corresponding pseudocode (\Cref{fig:rdfsbase:glm:pseudo}).
Annotations have the same meaning as described in \S\ref{sec:uspectre:annotations}. %
\texttt{r64} is the architectural output register of the \texttt{rdfsbase}/\texttt{rdgsbase} instruction.

If \texttt{CR4.FSGSBASE} is reset, \texttt{rdfsbase} should raise a general protection fault (\#GP)~\cite{intel-sdm}.
However, the processor mispredicts that the control-flow path raising \#GP is not-taken and instead transiently executes the rest of the microprogram and then subsequent attacker macro-ops as if \texttt{CR4.FSGSBASE} were set.
The transiently executed attacker macro-ops can then leak the value of the \texttt{FS}/\texttt{GS} base registers via a transmitter (\S\ref{sec:background:hw-side-channels}).
We name this attack \meb{}-FSGSBASE.

\begin{figure}
\begin{subfigure}{\linewidth}
\begin{lstlisting}[style=example, escapeinside=??]
xgetbv:
  ?\lstseq{tmp0 = RDCREG64(CR4)}?
  ?\lstbr{BT\_UJMPNC(tmp0, 0x12, \#UD)}?
  ?\lsttrans{tmp0 = SHR32(rcx, 1)}?
  ?\lsttrans{UJMPNZ(tmp0, \#GP)}??\label{line:xgetbv:br2}?
  ?\lsttrans{\hl{tmp0} = READURAM64(0x5b)}?
  ?\lsttrans{rdx = ZEROEXT32(0)}?
  ?\lsttrans{\hl{tmp0} = SHR64(\hl{tmp0}, 0x38)}?
  ?\lsttrans{\hl{rax} = OR64(\hl{tmp0}, 1)}?
  ?\lsttrans{SEQW UEND0}?
#UD:
  SIGEVENT(0x19)
  ...
#GP:
  SIGEVENT(0xf5)
  ...
\end{lstlisting}
\caption{MEB-vulnerable microprogram (simplified)}
\label{fig:xgetbv:glm:ucode}
\end{subfigure}

\vspace{5mm}

\begin{subfigure}{\linewidth}
\begin{lstlisting}[style=example, escapeinside=??]
?\lstbr{if (!CR4.OSXSAVE)}?
  raise #UD;
?\lsttrans{if (rcx > 1)}?
  raise #GP;
?\lsttrans{rdx = 0;}?
?\lsttrans{\hl{rax} = \hl{XCR0};}?
\end{lstlisting}
\caption{Pseudocode for vulnerable microprogram}
\label{fig:xgetbv:glm:pseudo}
\end{subfigure}

\vspace{5mm}

\begin{subfigure}{\linewidth}
\begin{lstlisting}[style=example, escapeinside=??]
// assume CR.OSXSAVE = 0
?\lstseq{u32 xcrl, xcrh;}?
?\lstbr{asm("xgetbv" : "=a"(\hl{xcrl}), "=d"(\hl{xcrh}) : "c"(0));}?
?\lsttrans{transmit(\hl{xcrl}); transmit(\hl{xcrh});}?
\end{lstlisting}
\caption{Rogue System Register Read attack (architectural program)}
\label{fig:xgetbv:glm:attack}
\end{subfigure}

\caption{MEB-vulnerable microprogram implementing \texttt{xgetbv} on Goldmont.}
\label{fig:xgetbv:glm}

\end{figure}

\paragraph{\meb{}-OSXSAVE (\texttt{xgetbv})}
\texttt{xgetbv}---``get value of extended control register''---copies the value in XCR0 into \texttt{rax} (and zeroes \texttt{rdx})~\cite{intel-sdm}.
However, \texttt{xgetbv} is only available if the \texttt{CR4.OSXSAVE} is set;
if reset, attempting to execute \texttt{xgetbv} raises an undefined opcode (\verb|#UD|) exception.
Goldmont's microprogram for \texttt{xgetbv} (\Cref{fig:xgetbv:glm}) uses a vulnerable conditional microcode branch to implement this check.
If \texttt{CR4.OSXSAVE=0}, it sequentially branches to an undefined opcode (\verb|#UD|) subroutine.
However, the branch is predicted not-taken and transiently executes the rest of the microprogram and subsequent macro-ops, which may leak the value of \texttt{XCR0} held in \texttt{rax}.

Per our naming convention, we call this attack \meb{}-OSXSAVE.
\texttt{xgetbv}'s microprogram contains a second conditional \ubranch{} (\cref{line:xgetbv:br2} in \Cref{fig:xgetbv:glm:ucode});
however, this branch does not introduce any additional vulnerabilities, since it cannot be exploited to return any additional privileged state beyond \texttt{XCR0}.

\begin{figure}

\begin{subfigure}{\linewidth}
\begin{lstlisting}[style=example, caption={Goldmont microprogram implementing \texttt{rdpmc} (simplified and without microcode updates)}, escapeinside={?}{?}, label=lst:rdpmc]
rdpmc:
  ?\textbf{tmp3 = ROL32(rcx, 2)}?
  ?\textbf{tmp1 = ADD8(tmp3, 0x2260)}?
  ?\textbf{\hl{tmp7} = RDCREG64(tmp1)}?
  ?\textbf{tmp1 = NOTAND32(}?
    ?\textbf{0x40000003, rcx)}?
  ?\textbf{tmp2 = RDCREG64(CR4)}?
  ?\textbf{tmp2 = AND32(0x100, tmp2)}?
  ?\textcolor{red}{\textbf{UJMPZ(tmp2, \#GP)}}?
  ?\textcolor{gray}{\textbf{UJMPNZ(tmp1, \#GP)}}?
  ?\textcolor{gray}{\textbf{CMP\_UJMPZ(tmp3, 13, \#GP)}}?
  ?\textcolor{gray}{\textbf{\hl{rax} = ZEROEXT32(\hl{tmp7})}}?
  ?\textcolor{gray}{\textbf{\hl{rdx} = SHR\_DSZ64(\hl{tmp7}, 32)}}?
  ?\textcolor{gray}{\textbf{SEQW UEND0}}?

#UD:
  SIGEVENT(0x19)
  ...

  
\end{lstlisting}
\caption{MEB-vulnerable microprogram (simplified)}
\label{fig:rdpmc:glm:ucode}
\end{subfigure}

\vspace{5mm}

\begin{subfigure}{\linewidth}
\begin{lstlisting}[style=example, caption={Pseudocode for \texttt{rdpmc}}, label=lst:rdpmc:pseudo, escapeinside={?}{?}]
?\lstseq{pmc = rol(rcx, 2);}?
?\lstseq{addr = pmc + 0x2260;}?
?\lstseq{\hl{data} = read\_creg(addr);}?
?\lstbr{if (!CR4.PCE)}?
  raise #GP;
?\lsttrans{if ((rcx \& ~0x40000003))}?
  raise #GP;
?\lsttrans{if (pmc == 13)}?
  raise #GP;
?\lsttrans{\hl{rax} = \hl{data};}?
?\lsttrans{\hl{rdx} = \hl{data} >{}> 32;}?
\end{lstlisting}
\caption{Pseudocode for vulnerable microprogram}
\label{fig:rdpmc:glm:pseudo}
\end{subfigure}

\vspace{5mm}

\begin{subfigure}{\linewidth}
\begin{lstlisting}[style=example, escapeinside=??]
// assume CR.PCE = 0
?\lstseq{u32 pmch, pmcl;}?
?\lstbr{asm("rdpmc" : "=a"(\hl{pmcl}), "=d"(\hl{pmch}) : "c"(0));}?
?\lsttrans{transmit(\hl{pmch}); transmit(\hl{pmcl});}?
\end{lstlisting}
\caption{Rogue System Register Read attack (architectural program)}
\label{fig:rdpmc:glm:attack}
\end{subfigure}

\caption{MEB-vulnerable microprogram implementing \texttt{rdpmc} on Goldmont.}
\label{fig:rdpmc:glm}

\end{figure}

\paragraph{Example: \texttt{rdpmc}}
\texttt{rdpmc}---``read performance-monitoring counters''---reads the 64-bit performance counter at index \texttt{rcx} into output registers \texttt{edx:eax}.
However, \texttt{rdpmc} is only available if \texttt{CR4.PCE} is set; otherwise, it raises a general protection (\verb|#GP|) exception.
Below is the vulnerable microprogram for \texttt{rdpmc}, without microcode patches applied. (Intel's microcode patch fixes an unrelated vulnerability in \texttt{rdpmc}, which we discuss in \S\ref{sec:new:pcidx}.)

Notably, \texttt{rdpmc}'s microprogram contains \textit{three} vulnerable microcode branches.
Only the \textit{first} branch, however, gives rise to \texttt{rdpmc}'s vulnerability to Rogue System Register Read:
it checks if the bit \texttt{CR4.PCE} is set, and if not, branches to code that raises a \verb|#GP| exception.
However, it is statically predicted not-taken, even if \texttt{CR4.PCE=0}, allowing the following micro-ops to transiently copy the requested performance counter value into output registers \texttt{rax} and \texttt{rdx}, which a malicious program can then transiently leak.
Per our naming convention, we call this attack \meb{}-PCE.

The two following conditional branches give rise to a \textit{separate vulnerability} that, to the best of our knowledge, has \textit{never been publicly disclosed} but has been silently fixed by Intel in a microcode update.
We discuss this novel vulnerability in \S\ref{sec:new:pcidx} and present Intel's patch to \texttt{rdpmc}'s microprogram in \Cref{fig:uslh:rdpmc}.

\subsubsection{Rogue System Register Read on other Intel Microarchitectures}
\label{sec:existing:rsrr:other}
In \S\ref{sec:existing:rsrr:goldmont}, we show that all three Rogue System Register Read vulnerabilities on Goldmont are in fact due to microcode exception bypasses.
However, according to Intel~\cite{intel-rogue-system-register-read-instructions}, many other instructions---such as \texttt{rdtsc}, \texttt{swapgs}, \texttt{smsw}---are vulnerable to Rogue System Register Read on some Intel processors.
We have not yet tested these other vulnerable instructions for microcode exception bypass. %

\subsection{Meltdown-BND (\meb{}-BND)} %
\label{sec:existing:bnd}
We report for the first time that Meltdown-BND %
is a \textit{\meb{} vulnerability}, not a Meltdown variant involving delayed fault handling as originally reported~\cite{canella-systematic}.
However, we empirically confirm that Canella et al.'s classification of a related vulnerability, Meltdown-MPX, is indeed an instance of Meltdown involving delayed fault handling.

\subsubsection{Background}
Meltdown-BND and Meltdown-MPX, first reported by Canella et al.~\cite{canella-systematic}, exploit transient execution following a failed bounds check performed by a hardware bounds check instruction. %
Architecturally, the bounds check instruction raises a bounds-range-exceeded exception (\texttt{\#BR}) when the pointer or array index passed to it is out of bounds. 
Before the exception is raised, however, the CPU transiently executes subsequent instructions, which can access and leak out-of-bounds sensitive data.
Meltdown-BND exploits transient execution following a faulting \texttt{bound} instruction (which is not available in 64-bit mode)~\cite{intel-sdm}.
Meltdown-MPX exploits transient execution following a faulting \texttt{bndcl}, \texttt{bndcu}, or \texttt{bndcn} instruction, which are provided as part of Intel's (now deprecated) Memory Protection Extensions (Intel MPX)~\cite{intel-sdm}.

Canella et al. concluded that both Meltdown-BND and Meltdown-MPX arise due to Meltdown-style lazy handling of the \texttt{\#BR} exception.
To evaluate these claims, we perform a case study on the Goldmont microarchitecture to better understand the microprograms that compromise the vulnerable instructions in Meltdown-BND and Meltdown-MPX.

\begin{figure}

\begin{subfigure}{\linewidth}
\begin{lstlisting}[style=example, escapeinside=??]
bound r32, m32:
  ?\textbf{tmp0 = LD32(DS, m32base, m32idx, m32scale,}?
              ?\textbf{m32disp)}?
  ?\textbf{tmp1 = LA2LIN32(DS, m32base, m32idx,}?
                  ?\textbf{m32scale, m32disp)}?
  ?\textbf{tmp0 = SUB32(tmp0, r32)}?
  ?\textcolor{red}{\textbf{UJMPL(tmp0, \#BR)}}?
  ?\textcolor{gray}{\textbf{tmp0 = LD32(tmp1, 4)}}?
  ?\textcolor{gray}{\textbf{tmp0 = SUB32(tmp0, r32)}}?
  ?\textcolor{red}{\textbf{UJMPG(tmp0, \#BR)}}?
  ?\textcolor{gray}{\textbf{NOP}}?
  ?\textcolor{gray}{\textbf{SEQW UEND0}}?

#BR:
  tmp1 = RDCREG64(0x701)
  ...
  SIGEVENT(0x15)
  ...
\end{lstlisting}
\caption{MEB-vulnerable microprogram (simplified)}
\label{fig:bound:glm:ucode}
\end{subfigure}

\vspace{5mm}

\begin{subfigure}{\linewidth}
\begin{lstlisting}[style=example, escapeinside=??]
void bound(i32 r32, i32 *m32) {
  ?\lstseq{i32 lower\_bound = m32[0];}?
  ?\lstbr{if (r32 < lower\_bound)}?
    raise #BR;
  ?\lsttrans{i32 upper\_bound = m32[1];}?
  ?\lstbr{if (r32 > upper\_bound)}?
    raise #BR;
}
\end{lstlisting}
\caption{Pseudocode for vulnerable microprogram}
\label{fig:bound:glm:pseudo}
\end{subfigure}

\vspace{5mm}

\begin{subfigure}{\linewidth}
\begin{lstlisting}[style=example, escapeinside=??]
void foo(i32 idx, i32 bnd[2], u8 *buf) {
  // assume idx > bnd[1]
  ?\lstbr{asm ("bound \%0, \%1" :: "r"(idx), "m"(bnd));}?
  ?\lsttrans{u8 \hl{value} = buf[idx];}?
  ?\lsttrans{transmit(\hl{value});}?
}
\end{lstlisting}
\caption{Meltdown-BND attack (architectural program)}
\label{fig:bound:glm:attack}
\end{subfigure}

\caption{MEB-vulnerable microprogram implementing \texttt{bound} on Goldmont.}
\label{fig:bound:glm}

\end{figure}

\subsubsection{Case Study: \meb{}-BND (Meltdown-BND) on Intel Goldmont}
We find that Meltdown-BND, which involves transient execution following a faulting \texttt{bound} instruction, is a \ourspectre{} microcode exception bypass vulnerability. Thus, we refer to it as \textit{\meb{}-BND}.

Below is the microprogram implementing \texttt{bound r32, m32} on Goldmont,
where \texttt{ridx} is a 32-bit array index being checked against a 32-bit lower and upper bound stored at \texttt{*m32} and \texttt{*(m32+4)}, respectively.

Goldmont's microprogram implementing \texttt{bound} contains two vulnerable conditional branches,
the first checking the lower bound and the second checking the upper bound.
Both branch to microcode that raises a \verb|#BR| exception if the check fails.
However, since they are statically mispredicted not-taken (Observation 1, \S\ref{sec:ubranches:obs1}), they allow subsequent macro-ops to execute even if either check fails.
We conclude that the vulnerability in \texttt{bound}
is in fact an instance of \meb{}, not Meltdown.

\begin{figure}
\begin{subfigure}{\linewidth}
\begin{lstlisting}[style=example, escapeinside=??]
bndcn bnd, m64:
  ?\textbf{tmp0 = LD64(m64base, m64idx, m64scale, m64disp)}?
  ?\textbf{tmp0 = SUB64(bnd.ub, tmp0)}?
  ?\textbf{tmp0 = SELECTBE(tmp0, 0x15)}?
  ?\textbf{SIGEVENT(tmp0)}?
  ?\textbf{SEQW UEND0}?
\end{lstlisting}
\caption{Non-\uspectre{}-vulnerable microprogram}
\label{fig:bndcn:glm:ucode}
\end{subfigure}

\vspace{5mm}

\begin{subfigure}{\linewidth}
\begin{lstlisting}[style=example]
void *ptr = *m64;
uint64_t fault = (ptr > bnd.ub ? #BR : 0);
raise fault; // if fault == 0, then no-op
\end{lstlisting}
\caption{Pseudocode for non-vulnerable microprogram}
\label{fig:bndcn:glm:pseudo}
\end{subfigure}
\caption{Non-\uspectre{}-vulnerable microcode implementing \texttt{bndcn} on Goldmont, which gives rise to Meltdown variant Meltdown-MPX~\cite{canella-systematic}.}
\end{figure}

\subsubsection{Case Study: Meltdown-MPX on Intel Goldmont}
We confirm Canella et al.'s classification of Meltdown-MPX as a Meltdown variant~\cite{canella-systematic} by examining the MPX bounds check instruction \texttt{bndcn} on Goldmont.
We choose \texttt{bndcn} rather than \texttt{bndcl} or \texttt{bndcu} because it is the only microcoded instruction of the three.
This means that we can view its microprogram implementation in the MSROM (\S\ref{sec:background:msrom}).

\Cref{fig:bound:glm} contains the microprogram implementing \texttt{bndcn} on Goldmont.
The instruction \texttt{bndcn bnd, m64} raises a \verb|#BR| exception if the 64-bit address pointed to by \texttt{m64} is greater than the upper bound \texttt{bnd.UB} provided in bounds register \texttt{bnd}, where \texttt{bnd.UB} is not in one's complement form~\cite{intel-sdm}.

Notably, \Cref{fig:bndcn:glm:ucode} contains no conditional \ubranches{}.
However, Canella et al. still observe transient execution following these instructions;
thus, we concur that Meltdown-MPX is indeed due to delayed fault handling---in this case, the \verb|#BR| raised by the fourth micro-op \texttt{SIGEVENT(tmp0)} in \Cref{fig:bndcn:glm:ucode}.

\subsection{Zero Dividend Injection (\mvi{}-ZDI)}
\label{sec:existing:zdi}
We report for the first time that Zero Dividend Injection (ZDI)~\cite{revizor++} occurs due to a \textit{mispredicted conditional \ubranch{}}
and thus is an instance of microcode value injection (\S\ref{sec:uspectre:mvi}).
We call the resulting vulnerability \textit{\mvi{}-ZDI}.

\paragraph{Background}
Oleksenko et al.~\cite{revizor++} recently discovered that while performing 64-bit division, some Intel processors speculate that the upper 64 bits of the 128-bit dividend are zero.
They call this \textit{zero dividend injection (ZDI)}, which occurs as follows.
The division instruction \texttt{div r64} architecturally computes the division \texttt{rdx:rax / r64}.
However, they observe that if \texttt{rdx} is non-zero and takes a long time to resolve (e.g., due to a cache miss),
then the processor transiently performs the division \texttt{0:rax / r64} (as if \texttt{rdx=0}) and then transiently executes subsequent instructions with the incorrect quotient (in \texttt{rax}) and remainder (in \texttt{rdx}).

Oleksenko et al. attribute this transient behavior to ``the first documented case of value prediction in a commercial CPU''~\cite{revizor++}.
Contrary to their conclusion, we deduce that ZDI is in fact an instance of \mvi{} by performing the following case study on a Haswell CPU.

\begin{figure}
\begin{lstlisting}[style=example, escapeinside=??]
div r64:
  ?\textbf{...}?
  ?\textbf{tmp0 = OR64(rdx, rdx)}?
  ?\textcolor{red}{\textbf{UJMPNZ(tmp0, .slow)}}?
?\textcolor{gray}{\textbf{.fast:}}?
  ?\textcolor{gray}{\textbf{... // perform 64-bit by 64-bit division}}?
  ?\textcolor{gray}{\textbf{rax = ... // quotient}}?
  ?\textcolor{gray}{\textbf{rdx = ... // remainder}}?
  ?\textcolor{gray}{\textbf{SEQW UEND0}}?

.slow:
  ... // perform 128-bit by 64-bit division
  rax = ... // quotient
  rdx = ... // remainder
  SEQW UEND0
}
\end{lstlisting}

\caption{Hypothesized MVI-vulnerable microprogram for Haswell's \texttt{div} microprogram, giving rise to \mvi{}-ZDI.}
\label{fig:div:hsw}

\end{figure}

\paragraph{Case Study on Haswell}
We determined that ZDI is due to a mispredicted \ubranch{} as follows.
First, we confirmed that our Haswell core exhibits ZDI by leaking the transiently computed quotient in \texttt{rax} via a cache side channel.
No other cores we tested exhibited ZDI.
Second, since Haswell's microcode is not public, we ran a series of nanobenchmarks with nanoBench~\cite{nanobench} to profile the sequential and transient execution paths of the microprogram implementing 64-bit division. %
The first nanobenchmark, ``\texttt{mov rdx, 0; mov rax, 0; mov rcx, 2; div rcx}'' sets \texttt{rdx} to zero and issues/retires 39/39 micro-ops.
The second nanobenchmark, ``\texttt{mov rdx, 1; mov rax, 0; mov rcx, 2; div rcx}'' sets \texttt{rdx} to a nonzero value and issues/retires 61/49 micro-ops. 
In both cases, all micro-ops are delivered from the MSROM (\S\ref{sec:background:msrom}) according to the \verb|IDQ.MS_UOPS| performance counter~\cite{intel-sdm}, thus \texttt{div} is microcoded.

Because \texttt{div}'s microprogram \textit{retires} a different number of micro-ops depending on whether \texttt{rdx} is zero (39 vs. 49), we conclude that the microprogram must taken one of two different control-flow paths.
This implies the presence of a \textit{conditional \ubranch{}} that depends on the value of \texttt{rdx}, which holds the upper 64 bits of the dividend.
Furthermore, we can deduce that the branch was \textit{mispredicted} when $\texttt{rdx}=1$, since more instructions are issued than retired (61 vs. 49), i.e., $61-49=12$ instructions were transiently issued.

We conclude that the conditional \ubranch{}'s predicate must be ``$\texttt{rdx} \neq 0$,'' since \ubranches{} are statically (mis)predicted not-taken (Observation 1, \S\ref{sec:ubranches:obs1}).
Thus, we infer that the microprogram implementing \texttt{div} has the structure shown in \Cref{fig:div:hsw}.
We conclude that ZDI is an instance of \mvi{}, since the \texttt{div}'s mispredicted conditional \ubranch{} transiently exposes an invalid quotient and remainder to subsequent macro-ops. 

\begin{figure}
\begin{subfigure}{\linewidth}
\begin{lstlisting}[style=example, escapeinside=??]
repne_scasb:
  ?\textbf{tmp4 = OR64(rcx)}?
  ?\textcolor{red}{\textbf{UJMPZ(tmp4, .zerolen)}}?
.loop:
  ?\lsttrans{\textbf{\hl{tmp1} = LDZXn(rdi)}}?
  ?\lsttrans{\textbf{\hl{tmp10} = SUB64(\hl{tmp1}, rax)}}?
  ?\lsttrans{\textbf{rdi = ADDSUB64(rdi, 1)}}?
  ?\lsttrans{\textbf{tmp4 = SUB64(tmp4, 1)}}?
  ?\lstbr{\textbf{UJMPZ(\hl{tmp10}, .done)}}? // transmitter
  ?\lstbr{\textbf{UJMPZ(tmp4, .done)}}?
  ?\lsttrans{\textbf{SEQW GOTO .loop}}?

.done:
  GENARITHFLAGS(0x3f, tmp10)
  rcx = ZEROEXT64(tmp4, rcx)
  SFENCE(0)
  SEQW SYNCWAIT
  SEQW UEND0

.zerolen:
  rcx = ZEROEXT64(rcx)
  SEQW UEND0
\end{lstlisting}
\caption{MIL-vulnerable microprogram (simplified)}
\label{fig:scasb:glm:ucode}
\end{subfigure}

\vspace{5mm}

\begin{subfigure}{\linewidth}
\begin{lstlisting}[style=example, escapeinside=??]
?\lstbr{if (rcx == 0)}?
  return;
?\lsttrans{while (1) \{}?
  ?\lsttrans{u8 \hl{b} = *rdi;}?
  ?\lsttrans{--rcx;}?
  ?\lsttrans{++rdi;}?
  ?\lstbr{if (\hl{b} == al)}? // transmitter
    return;
  ?\lstbr{if (rcx == 0)}?
    return;
}
\end{lstlisting}
\caption{Pseudocode for the vulnerable microprogram}
\label{fig:scasb:glm:pseudo}
\end{subfigure}

\caption{MIL-vulnerable microprogram implementing \texttt{repne scasb} on Goldmont, giving rise to \mil{}-SCO.}
\label{fig:scasb:glm}

\end{figure}

\subsection{String Comparison Overrun (\mil{}-SCO)}
\label{sec:existing:sco}
We are the first to report that String Comparison Overrun (SCO)~\cite{revizor++} is an \textit{instance of \mil{}}. %
Thus, we refer to these vulnerabilities as \mil{}-SCO.

\paragraph{Background}
Oleksenko et al. also discovered that some x86 string instructions---in particular \texttt{repe/repne cmps} and \texttt{repe/repne scas}---compare data past the given string length (in \texttt{rcx}) and can leak the outcome of the comparisons into the cache state~\cite{revizor++}.
They call this vulnerability \textit{String Comparison Overrun} (SCO).

\paragraph{Case Study on Goldmont}
We reproduce the (simplified) Goldmont microcode for the SCO-vulnerable macro-op \texttt{repne scasb} (scan string pointed to by \texttt{rdi} for byte in register \texttt{al}) in \Cref{fig:scasb:glm}.

The microprogram implementing \texttt{repne scasb} \textit{always} predicts that (1) the length is non-zero, even after being decremented, 
and (2) the current string byte does not equal the value in \texttt{al}.
Thus, \texttt{repne scasb} compares past the end of the string because its microprogram speculates in an infinite loop.
This speculative loop is only broken when a misprediction of one of the microcode branches is detected. %

Importantly, an attacker can \textit{control} how long it takes for the branch predicates to resolve by either (a) setting the string length in \texttt{rcx} to be result of an arbitrarily long latency computation or (b) flushing portions of the string pointed to by \texttt{rdi} from the cache to delay the resolution of arbitrary byte comparisons that may cause the loop to exit.

\section{New \Ourspectre{} Vulnerabilities}
\label{sec:new}

In this section, we present \textit{two novel \meb{}} vulnerabilities
and \textit{three new \mvi{} speculation primitives} that we have discovered.
In particular, the new vulnerability \meb{}-PCIDX (\S\ref{sec:new:pcidx}) demonstrates how \textit{uniquely severe} \uspectre{} vulnerabilities can be, with the potential to leak \textit{large swaths of microarchitectural state} that is not intended to be accessible to \textit{any} program running on the hardware, \textit{ever}. %

While we did not discover any new \mil{} attacks,
we show how inverting the taken/not-taken directions of a \textit{single} conditional \ubranch{} while preserving its semantics introduces a \mil{} vulnerability that enables an attacker to leak all of its physical page mappings (\S\ref{sec:new:0f0c}).

\begin{figure}
\begin{lstlisting}[style=example, escapeinside=??]
into:
  ?\textbf{tmp2 = MOVE\_MERGEFLAGS32(4)}?
  ?\textcolor{red}{\textbf{UJMPO(tmp2, \#OF)}}?
  ?\textcolor{gray}{\textbf{SEQW UEND0}}?
#OF:
  ...
\end{lstlisting}
\caption{MEB-vulnerable microprogram implementing \texttt{into} on Goldmont (simplified), giving rise to \meb{}-OF (\S\ref{sec:new:of}).}
\label{fig:into:glm}
\end{figure}

\subsection{Overflow Check Bypass (\Ourspectre{}-MEB-OF)}
\label{sec:new:of}
We present \textit{Overflow Check Bypass}, or \meb{}-OF, which is an instance of microcode exception bypass.
\meb{}-OF exploits a vulnerability in the microprogram that implements the \texttt{into} instruction,
which generates an overflow trap (\verb|#OF|) if the overflow flag (\texttt{OF}) is set.\footnote{\texttt{into} is not available in 64-bit mode.}
On Goldmont, we find that this check is implemented using a \textit{vulnerable conditional \ubranch{}} to a routine that raises the \verb|#OF| exception.
This branch is statically predicted not-taken (per Observation 1, \S\ref{sec:ubranches:obs1}) and thus allows transient execution of subsequent macro-ops despite an overflow having occurred.

An attacker can exploit \ourspectre{}-MEB-OF to transiently compute on invalid values that were the result of a signed integer overflow,
allowing an attacker to potentially access and leak out-of-bounds data.

\subsection{Out-of-Bounds Performance Counter Read (\meb{}-PCIDX)}
\label{sec:new:pcidx}
Recall from \Cref{lst:rdpmc} that on Goldmont, the microprogram implementing \texttt{rdpmc} contains \textit{three} vulnerable microcode branches.
The first branch is responsible for \ourspectre{}-MEB-PCE, which is an instance of the \meb{}-REG subclass of \uspectre{} attacks (\S\ref{sec:existing:rsrr}).
The \textit{second and third} branches, which implement bounds checks on the provided performance counter index, give rise to a \textit{new vulnerability}, 
which we call \textit{Out-of-Bounds Performance Counter Read}, or \meb{}-PCIDX for short.

\ourspectre{}-MEB-PCIDX allows an attacker to pass a malicious out-of-bounds performance counter index to \texttt{rdpmc} via architectural register \texttt{rcx}, which causes the microprogram to transiently read from an out-of-bounds control register address,
bypass all the bounds checks via microcode branch mispredictions, and transiently pass the out-of-bounds data to subsequent macro-ops which leak it.

\begin{figure}
\begin{subfigure}{\linewidth}
\begin{lstlisting}[style=example, escapeinside=??]
rdpmc:
  ?\textbf{tmp3 = ROL32(rcx, 2)}\label{line:rol}?
  ?\textbf{tmp1 = ADD8(tmp3, 0x2260)}\label{line:add}?
  ?\textbf{tmp7 = RDCREG64(tmp1)}\label{line:rdcreg}?
  ?\textbf{tmp1 = NOTAND32(}\label{line:notand1}?
    ?\textbf{0x40000003, rcx)}\label{line:notand2}?
  ?\textbf{tmp2 = RDCREG64(CR4)}?
  ?\textbf{tmp2 = AND32(0x100, tmp2)}?
  ?\textcolor{black}{\textbf{UJMPZ(tmp2, \#GP)}}?
  ?\textcolor{red}{\textbf{UJMPNZ(tmp1, \#GP)}}\label{line:bounds-check}? // bounds check
  ?\textcolor{red}{\textbf{CMP\_UJMPZ(tmp3, 13, \#GP)}}\label{line:hole-check}? // hole check
  ?\textcolor{gray}{\textbf{rax = tmp7}}\label{line:copy-rax}?
  ?\textcolor{gray}{\textbf{rdx = SHR\_DSZ64(tmp7, 32)}}\label{line:copy-rdx}?
  ?\textcolor{gray}{\textbf{SEQW UEND0}}?
#GP:
  SIGEVENT(0x19)
  ...
\end{lstlisting}
\caption{MEB-vulnerable microprogram implementing \texttt{rdpmc} on Goldmont (reproduced with emphasis on \ubranches{} that give rise to \meb{}-PCIDX).}
\label{fig:new:pcidx:ucode}
\end{subfigure}

\vspace{5mm}

\begin{subfigure}{\linewidth}
\begin{lstlisting}[style=example]
int junk;
char arr[256 * 512];
void leak_creg(u16 creg, u64 byte_sel) {
  // 0x2200 <= creg < 0x2300
  u32 phys_pmc = (creg - 0x2260) & 0xFF;
  u32 pmc = (phys_pmc >> 2) | (phys_pmc << 30);
  u32 eax, edx;
  asm ("rdpmc" : "=a"(eax), "=d"(edx) : "c"(pmc));
  // TRANSIENT EXECUTION WINDOW //
  u64 data = ((u64) edx << 32) | eax;
  u64 byte = (data >> (byte_sel * 8)) & 0xFF;
  junk &= arr[byte * 512];
  // END TRANSIENT EXEC. WINDOW //
}
\end{lstlisting}
\caption{Proof-of-concept attack exploiting \meb{}-PCIDX on Goldmont (architectural program).}
\label{fig:new:pcidx:attack}
\end{subfigure}

\caption{\ourspectre{}-MEB-PCIDX vulnerability in the microprogram implementing \texttt{rdpmc} on Goldmont.}
\label{fig:new:pcidx}
\end{figure}

\subsubsection{Detailed Explanation and Analysis}
We reproduce \texttt{rdpmc}'s microprogram in \Cref{fig:new:pcidx:ucode}, 
this time annotated assuming the \texttt{rdpmc} instruction is enabled, so that the first branch is not mispredicted.
An attacker can coerce \texttt{rdpmc}'s microprogram to transiently return out-of-bounds control register data to subsequent macro-ops as follows.

\textit{Reading from out-of-bounds control registers:}
\Cref{line:rol} in \Cref{fig:new:pcidx:ucode} computes the physical performance counter index from the logical index provided in architectural register \texttt{rcx} and assigns it to scratch register \texttt{tmp3}.
To compute the address of the control register holding the performance counter data corresponding to the physical index, 
\cref{line:add} indexes into the performance counter array based at \texttt{0x2260} using the physical performance counter index in \texttt{tmp3}.
Interestingly, this is done using an 8-bit addition with a 16-bit operand via the \texttt{ADD8} micro-op. 
We experimentally determined the semantics of the mixed-bitwidth addition to be the following.
First, the \texttt{ADD8} performs a conventional 8-bit addition between the least-significant bytes of \texttt{0x2260} and \texttt{tmp3}, wrapping around on overflow.
Then, it extends the 8-bit sum with the upper 8 bits of the immediate 0x2260 and writes the result into \verb|tmp1|.
Thus, \cref{line:add} effectively computes \verb|tmp1 = (tmp3 + 0x60) % 256 + 0x2200|.
Since the attacker fully controls \texttt{tmp3} via the input \texttt{rcx},
the attacker can choose \texttt{rcx} such that \cref{line:add} computes an \textit{arbitrary control register address in the range 0x2200-0x2300}.
\Cref{line:rdcreg} then reads the data at this control register address into \texttt{tmp7}.

\textit{Mispredicting the bounds checks:}
\Cref{line:notand1,line:notand2,line:bounds-check,line:hole-check} check the validity of the attacker-controlled performance counter index passed in \texttt{rcx}.
Specifically, if \texttt{rcx} is not in the range 0x0--0x3 or 0x40000000--0x40000002 (inclusive), then the conditional \ubranches{} on \cref{line:bounds-check,line:hole-check} are taken in order to raise a \verb|#GP| exception.
However, both branches are statically predicted not-taken, so the rest of the microprogram executes as if \texttt{rcx} were in bounds.

\textit{Exposing out-of-bounds control register data to subsequent macro-ops:}
\Cref{line:copy-rax,line:copy-rdx} copy the data read from the out-of-bounds control register into architectural registers \texttt{eax} (lower 32 bits) and \texttt{edx} (upper 32 bits).
Line 14 ends the microprogram, causing subsequent attacker macro-ops to execute.
Finally, the attacker can transiently leak the control register data in \texttt{eax} and \texttt{edx} using a side channel like the cache, as described below.

\subsubsection{Proof-of-Concept Exploit}
\Cref{fig:new:pcidx:attack} contains a proof-of-concept attack exploiting \meb{}-PCIDX to leak the values of arbitrary control registers in the range 0x2200--0x2300.
Using this approach on an Intel Goldmont core with no microcode updates applied, we were able to leak all 256 control registers in the range.
While most were zero (indicating that they are unimplemented), we found many control registers that held non-zero data.
We have not analyzed the purpose of control registers (aside from performance counter registers) in this range.

\subsubsection{Intel's Mitigation}
Intel appears to have silently patched the \ourspectre{}-MEB-PCIDX vulnerability years ago (we have not yet determined exactly when),
so any Goldmont-based systems that have applied Intel's most recent microcode updates are not vulnerable to \ourspectre{}-MEB-PCIDX.
We discovered this fix by comparing the Goldmont microcode with and without microcode patches using uCodeDisasm~\cite{hackers-ucodedisasm}.
We show Intel's fix in \Cref{fig:uslh:rdpmc} in \S\ref{sec:uslh:pcidx}, which inspires our \uspectre{} mitigation technique, \ourmitigation{} (\S\ref{sec:uslh}).
We have not been able to reproduce \meb{}-PCIDX on any Intel microarchitectures other than Goldmont.

\subsubsection{Implications}
Our \Ourspectre{}-MEB-PCIDX attack represents a \textit{new leakage scope} that, to our knowledge, has not been seen before in any transient execution attack:
the leakage of microarchitectural registers that are \textit{never} architecturally accessible and do not correspond to \textit{any} architectural state whatsoever, to our knowledge.
Neither Spectre nor Meltdown have been demonstrated to be capable of this.

However, given that (1) \ourspectre{}-MEB-PCIDX is capable of leaking control registers only in a restricted range 
and that (2) Intel has already fixed it, one might assume that \ourspectre{}-MEC-PCIDX is an interesting but low severity vulnerability that demonstrates the value of microcode updates.

Our perspective more pessimistic.
We see an extremely close shave where Intel was \textit{one micro-op away}---literally---from leaking the contents of \textit{all control registers} on the CRBUS,
which accounts for a large portion of microarchitectural state.
If an \texttt{ADD32} micro-op rather than a \texttt{ADD8} micro-op was used to compute the control register address in \cref{line:add} of \Cref{fig:new:pcidx:ucode},
then an attacker could coerce \texttt{rdpmc} to transiently return the data contained in \textit{any} control register in the range \texttt{0x0--0x100000000}, not just \texttt{0x2200--0x2300}.
This would potentially \textit{fully compromise the system's privacy and security}, 
depending on how much sensitive data is accessible via control registers
(our hypothesis is, \textit{a lot}).

Even worse, Intel microcode for microarchitectures other than Goldmont is not public.
Even if we can empirically verify that \texttt{rdpmc} is not vulnerable to \ourspectre{}-MEB-PCIDX on other microarchitectures,
there may be similar, yet-to-be-discovered \ourspectre{} vulnerabilities that allow full transient access to microarchitectural arrays.
We cannot know for sure unless Intel (and AMD) publish their microcode for all microarchitectures.

\subsection{New Instances of \Ourspectre{}-MVI}
\label{sec:new:mvi}
We discovered three new instances of microcode value injection (\mvi{}) on Goldmont,
arising due to \ubranch{} mispredictions in the microprograms of \texttt{cld}, \texttt{arpl}, and \texttt{enter}.
While we have observed subsequent macro-ops to transiently compute on the invalid outputs of these microprograms, 
we have so for not been able to demonstrate attacks exploiting this fact to transiently leak sensitive data.
However, we believe it likely that, under the right circumstances, these three vulnerable microprograms can enable subsequent macro-ops to transiently leak sensitive data.

\paragraph{\texttt{cld}}
\texttt{cld} architecturally clears the direction flag (\texttt{DF}).
However, \texttt{cld}'s microprogram always predicts that \texttt{DF=0} and thus does not update it. %
While we find that macro-ops that depend on \texttt{DF}, like \texttt{lodsb}, can transiently execute with \texttt{DF} incorrectly being set,
we have not yet been able to demonstrate leakage of transiently accessed data following such macro-ops.

\paragraph{\texttt{arpl}}
This instruction adjusts the 2-bit requestor privilege level (RPL) field of a 16-bit segment selector.
\texttt{arpl}'s microprogram predicts that the RPL field does not require adjustment and transiently executes subsequent macro-ops with the original RPL value.
We have not attempted a proof-of-concept for \texttt{arpl} yet.

\paragraph{\texttt{enter}}
The instruction \texttt{enter imm16, imm8} creates a \texttt{imm16}-byte stack frame for the procedure~\cite{intel-sdm}.
If \texttt{imm8} is zero, it is effectively the same as \texttt{push rbp; mov rbp, rsp; sub rsp, imm8}.
If \texttt{imm8} is non-zero, it pushes additional nested frame pointers onto the stack.
We find that the microprogram implementing \texttt{enter} statically predicts that \texttt{imm8} is zero, i.e., no nested frame pointers are pushed, regardless of the value of \texttt{imm8}.
This is remarkable, since the microcode branch checking whether \texttt{imm8} is zero does not depend on runtime data, only directly-encoded immediate fields that should be available at decode-time.
We have not attempted a proof-of-concept demonstrating leakage due to \texttt{enter}'s microcode value injection yet.

\subsection{Rogue Physical Address Load (\mil{}-PHYS, Hypothetical)}
\label{sec:new:0f0c}

During our analysis of Goldmont's public microcode,
we discovered a new undocumented instruction with opcode \texttt{0f 0c}.
This instruction maps to an intriguing microprogram with a complex control-flow graph that contains physical loads and stores to and from architectural register \texttt{rax} and performs port I/O, among other things. %
We have not yet determined what the instruction does, though we suspect it might involve some kind of inter-core communication.

Like the \texttt{udbgrd} and \texttt{udbgwr} undocumented instructions (with opcodes \texttt{0f 0e} and \texttt{0f 0f}) discovered by Ermolov et al.~\cite{hackers-writeup},
we find that \texttt{0f 0c}'s microprogram raises an undefined opcode (\verb|#UD|) exception unless the processor has been red-unlocked.
However, because the microprogram checks whether the instruction is enabled using a conditional branch to a \verb|#UD| routine, we observe transient execution along the not-taken path, even if the instruction is disabled.
Fortunately, we have verified through manual inspection that the not-taken direction of another branch falls through to the \verb|#UD| routine, so that ultimately no sensitive instructions are transiently executed.

\paragraph{Hypothetical vulnerability}
Had the taken/not-taken paths of one particular branch been switched, we believe that \textit{physical}-address loads and stores directly from register \texttt{rax}, which bypass address translation, could transiently execute, regardless of the privilege level.
This would have introduced a severe \ourspectre{}-MIL vulnerability, which an unprivileged attacker could exploit to infer its page table mappings from userspace.
We call this hypothetical vulnerability \textit{Rogue Physical Address Load}, or \mil{}-PHYS for short.

For example, an attacker could launch a 
Flush+Reload attack~\cite{flush+reload} to determine whether virtual page address $v$ maps to phyiscal page address $p$ as follows.
First, the attacker flushes $v$ from the cache.
Then, the attacker sets $\texttt{rax} \gets p$ and executes \texttt{0f 0c}, which architecturally raises a \verb|#UD| exception but transiently accesses $p$.
Finally, the attacker probes whether virtual address $v$ is now cached by timing the latency of a load from $v$.
If $v$ is cached, then $v$ maps to $p$.

\paragraph{Impact}
Although it appears that Goldmont's microprogram for the instruction \texttt{0f 0c} is not vulnerable to Rogue Physical Address Load,
we have inferred using microcode-related performance counters that \texttt{0f 0c} is implemented on \textit{all other Intel processors we evaluated}.
Since other Intel processors' microcode is not public, we cannot know whether \texttt{0f 0c} on those processors is vulnerable to Rogue Physical Address Load or related vulnerabilities. %
Even if all implementations of \texttt{0f 0c} on all Intel processors are secure,
there may be other undocumented instructions that transiently execute unsafe or leaky instructions before architecturally raising an undefined opcode exception.

\section{\ourslh{}: A \ourspectre{} Mitigation}
\label{sec:uslh}

We present \ourslh{}, a microcode defense against \ourspectre{}. %
\ourslh{} is inspired by speculative load hardening (SLH)~\cite{llvm-slh, ultimate-slh, spectre-declassified}, a Spectre-PHT defense, and Intel's microcode patch that fixes \meb{}-PCIDX (\Cref{fig:uslh:rdpmc}).
\ourslh{} can be deployed via microcode updates to patch \uspectre{} vulnerabilities in existing hardware and can be incorporated into directly the microcode of future hardware.

\ourslh{} works by using branchless logic to zero out any registers that may hold invalid or privileged data following a mispredicted \ubranch{}.
Specifically, sequentially following each conditional \ubranch{}, \ourslh{} inserts conditional select micro-op(s) (\texttt{SELECTcc}), predicated on the \textit{inverted condition} of the preceding \ubranch{}.
In effect, this conditional select zeroes out the register holding ``bad'' data unless if the prior branch's predicate resolved to \textit{false} (i.e., the static prediction of not-taken was correct).

\ourslh{} \textit{fully mitigates} all \ourspectre{}-MEB vulnerabilities that involve passing privileged data to subsequent macro-ops (e.g., all Rogue System Register Read vulnerabilities, \S\ref{sec:existing:rsrr}).
\ourslh{} \textit{partially mitigates} all \ourspectre{}-MVI attacks that involve directly passing invalid results to subsequent macro-ops (e.g., \mvi{}-ZDI, \S\ref{sec:existing:zdi}).
While \ourslh{} could theoretically mitigate \ourspectre{}-MIL attacks (e.g., \ourspectre{}-MIL-SCO), we predict the overhead would be too high due to the frequent occurence of tightly nested microcode loops.
Thus, other mitigations are needed for such \mil{} attacks.

\paragraph{Semantics of \texttt{SELECTcc}}
\ourslh{} relies on the conditional select micro-op, \texttt{SELECTcc}.
The micro-op \texttt{r0 = SELECTcc(r1, r2)} is equivalent to the C ternary expression \texttt{r0 = cc(r1.\allowbreak{}flags) ? r2 : 0}.
That is, if the flags associated with register \texttt{r1} satisfy condition code \texttt{cc}, then the data in \texttt{r2} is copied into output register \texttt{r0}; 
otherwise, \texttt{r0} is zeroed.

\subsection{Defending against \ourspectre{}-MEB with \ourslh{}}
\ourslh{} fully mitigates all \ourspectre{}-MEB vulnerabilities that involve passing privileged data to subsequent macro-ops,
including all \meb{}-REG (i.e., Rogue System Register Read) vulnerabilities (\S\ref{sec:existing:rsrr}) %
as well as \ourspectre{}-MEB-PCIDX (\S\ref{sec:new:pcidx}).

\begin{figure}
\begin{lstlisting}[style=example, escapeinside=??]
rdpmc:
  ?\lstseq{tmp3 = ROL32(rcx, 2)}?
  ?\lstseq{tmp1 = ADD8(tmp3, 0x2260)}?
  ?\lstseq{\lstsec{tmp7} = RDCREG64(tmp1)}?
  ?\lstseq{tmp1 = NOTAND32(}?
    ?\lstseq{0x40000003, rcx)}?
  ?\lstseq{tmp2 = RDCREG64(CR4)}?
  ?\lstseq{tmp2 = AND32(0x100, tmp2)}?
  ?\lstseq{UJMPZ(tmp2, \#GP)}?
  ?\lstbr{UJMP\underline{NZ}(tmp1, \#GP)}? // bounds check
  ?\lstbr{CMP\_UJMP\underline{Z}(tmp3, 13, \#GP)}? // hole check
  ?\lstnew{\lstsec{tmp7} = SELECT\underline{Z}(tmp1, \lstsec{tmp7})}\label{line:intels-mitigation}? // Intel's mitigation
  ?\lstnew{tmp7 = SELECT\underline{NZ}(tmp3, \lstsec{tmp7})}\label{line:our-mitigation}? // Our extension
  ?\lsttrans{\textbf{rax = tmp7}}?
  ?\lsttrans{\textbf{rdx = SHR\_DSZ64(tmp7, 32)}}?
  ?\lsttrans{\textbf{SEQW UEND0}}?

#GP:
  SIGEVENT(0x19)
  ...
\end{lstlisting}
\caption{Intel's mitigation of \ourspectre{}-MEB-PCIDX on Goldmont, plus our extension}
\label{fig:uslh:rdpmc}
\end{figure}

\subsubsection{Mitigating \ourspectre{}-MEB-PCIDX with \ourslh{}}
\label{sec:uslh:pcidx}
\textmu\-SLH is inspired by Intel's mitigation of \ourspectre{}-MEB-PCIDX (released via microcode update), shown in \Cref{fig:uslh:rdpmc}, along with our proposed extension.
We mark the new instructions inserted for both defenses in green.
Intel's microcode update inserts the first \texttt{SELECTcc} instruction (\cref{line:intels-mitigation}) to zero out the control register data in \texttt{tmp7} if the \ubranch{} implementing the bounds check is mispredicted not-taken.
As a result, if the performance counter index is out of bounds, \texttt{rax} and \texttt{rdx} are (transiently) zeroed, blocking any subsequent leakage.
However, if the performance counter index is in bounds but equals 13, the hole check (\cref{line:hole-check}) is still vulnerable and can transiently pass the corresponding control register's data to subsequent macro-ops.
To prevent this, we can simply insert a second conditional select directly following the second one (\cref{line:our-mitigation}).\footnote{In practice, the control register corresponding to performance counter index 13 is likely not implemented or, if it is, does not contain sensitive information. We include this second conditional select for completeness and to demonstrate the \ourslh{} approach.}

\begin{figure}
\begin{lstlisting}[style=example, escapeinside=??]
rdfsbase/rdgsbase r64:
  ?\lstseq{tmp0 = RDCREG64(CR4)}?
  ?\lstbr{BT\_UJMP\underline{NC}(tmp0, 16, \#GP)}?
  ?\lsttrans{\lstsec{tmp2} = RDSEG(FS/GS, BASE)}?
  ?\lstnew{tmp2 = SELECT\underline{C}(tmp0, \lstsec{tmp2})}?
  ?\lsttrans{r64 = ZEROEXTN(tmp2)}?
  ?\lsttrans{SEQW UEND0}?

#GP:
  SIGEVENT(0x19)
  ...
\end{lstlisting}
\caption{Our proposed \ourslh{} mitigation for \ourspectre{}-MEB-SEGBASE}
\label{fig:uslh:rdfsbase}
\end{figure}

\subsubsection{Mitigating Rogue System Register Read with \ourslh{}}
We believe that \textit{all Rogue System Register Read} vulnerabilities can be mitigated with \ourslh{}, 
based on our conclusion in \S\ref{sec:existing:rsrr:other} that all such vulnerabilities are likely due to mispredicted microcode branches (at least on Intel processors).
\Cref{fig:uslh:rdfsbase} shows how one would mitigate \ourspectre{}-MEB-SEGBASE with \ourslh{}.

We believe all other Rogue System Register Reads on all Intel processors can be mitigated similarly.
This begs the question: if mitigating Rogue System Register Read is so easy, why has Intel not mitigated it?
We consider two possibilities.

First, there are a limited of number of microcode patch registers (prior work finds that there are 32 on Goldmont~\cite{hackers-ucodedump})
but many Rogue System Register Read vulnerabilities (Intel's documentation lists 13 vulnerable instructions~\cite{intel-rogue-system-register-read-instructions}).
If Intel fixed all the Rogue System Register Read vulnerabilities with microcode patches, there might not be many patch registers left for more severe vulnerabilities (e.g., the worst-case scenario we describe in \S\ref{sec:new:pcidx}).
We speculate that Intel might be reserving the remaining microcode patch registers for higher-severity vulnerabilities.

Second, there might be a fundamental limitation of microcode updates that we are overlooking.
In our experiments in writing custom microcode for a red-unlocked Goldmont CPU~\cite{hackers-writeup},
we found that patching short microprograms like \texttt{rdfsbase}/\texttt{rdgsbase} would sometimes trigger faults and CPU freezes that we were unable to explain.
There may be restrictions on under what circumstances a microcode address can be patched, e.g., that it cannot be too close to the end of the microprogram. %
However, this is only speculation. %

\subsection{Mitigating against \ourspectre{}-MVI with \ourslh{}}
\ourslh{} partially mitigates all \mvi{} attacks that involve transiently passing invalid results to subsequent macro-ops, like in \mvi{}-ZDI.
Specifically, \ourslh{} zeroes out any invalid outputs produced by the microprogram following a misprediction microcode branch.
As a result, subsequent macro-ops can only transiently compute on one invalid value, zero.

This does not necessarily prevent all transient leaks in subsequent macro-ops, however.
For example, it is known that faulting loads transiently forwarding zero data to dependent micro-ops gives rise to LVI-Null attacks~\cite{lvi}.
The advantage of \ourslh{}, however, is that it architecturally exposes a universal transient zero-forwarding semantics to all instructions vulnerable to microcode value injection. %

\begin{figure}
\begin{lstlisting}[style=example, escapeinside=??]
div r64:
  ?\lstseq{...}?
  ?\lstseq{tmp0 = OR64(rdx, rdx)}?
  ?\lstbr{UJMP\underline{NZ}(tmp0, .slow)}?
?\lsttrans{.fast:}?
  ?\lsttrans{... // perform 64-bit by 64-bit division}?
  ?\lsttrans{\lstbad{rax} = ... // quotient}?
  ?\lstnew{rax = SELECT\underline{Z}(tmp0, \lstbad{rax})}?
  ?\lsttrans{\lstbad{rdx} = ... // remainder}?
  ?\lstnew{rdx = SELECT\underline{Z}(tmp0, \lstbad{rdx})}?
  ?\lsttrans{SEQW UEND0}?

.slow:
  ... // perform 128-bit by 64-bit division
  rax = ... // quotient
  rdx = ... // remainder
  SEQW UEND0
\end{lstlisting}
\caption{Our proposed \ourslh{} mitigation for \ourspectre{}-MVI-ZDI.}
\label{fig:uslh:div}
\end{figure}

\paragraph{Mitigating \ourspectre{}-MVI-ZDI}
\Cref{fig:uslh:div} shows how \ourslh{} mitigates \ourspectre{}-MVI-ZDI using two conditional selects at the end of the division's predicted path to conditionally zero out the quotient and remainder in \texttt{rax} and \texttt{rdx}, respectively, if the branch was mispredicted.
We highlight \lstbad{transiently invalid data} in red.

\subsection{Limitations of \ourslh{}}
In theory, \ourslh{} can mitigate \ourspectre{}-MIL vulnerabilities as well.
However, in practice, we suspect that the overhead of doing so would be too high and, instead, another mitigation approach should be used.
The only confirmed example of \ourspectre{}-MIL so far is \ourspectre{}-MIL-SCO, a vulnerability in the microprograms implementing \texttt{repe/repne cmps} and \texttt{repe/repne scas} (\S\ref{sec:existing:sco}).
Both microprograms on Goldmont contain tight loops (\Cref{fig:scasb:glm}), and \ourslh{} would require inserting \textit{two} conditional selects \textit{per iteration}.
For \texttt{repe/repne scas}, this would grow the loop body size from six micro-ops to eight micro-ops.
Thus, we would expect at least a 33\% increase in latency of the string operation, which is likely intolerable.

For now, we concur with Oleksenko et al.~\cite{revizor++} that the only practical defense against \ourspectre{}-MIL is to avoid using the vulnerable instructions altogether.

\section{Conclusion}
We present \uspectre{}, a novel class of transient execution attacks that exploit microcode branch mispredictions.
\uspectre{} is behind both long-known vulnerabilities, like Rogue System Register Read,
and also gives rise to entirely new kinds of vulnerabilities with unprecedented leakage potential, like Out-of-Bounds Performance Counter Read.
Unfortunately, Intel has not published its microcode for any of its microarchitectures.
As a result, we cannot know for sure what other \uspectre{} vulnerabilities lurk in the microprograms on modern x86 processors.

\begin{acks}
This work was supported in part by the National Science Foundation (NSF), under award numbers CNS-2153936 and CAREER CCF-2236855, and the Defense Advanced Research Projects Agency (DARPA) under contract W912CG-23-C-0025 and subcontract from Galois, Inc. We also gratefully acknowledge a gift from Intel.
\end{acks}

\nocite{speculative-buffer-overflows}

\bibliographystyle{plain}
\bibliography{references}

\end{document}